\documentclass[
prd,twocolumn,amsmath,amssymb]{revtex4}

\usepackage{graphicx}
\usepackage{amssymb}
\usepackage{amsmath}
\usepackage{dsfont}
\usepackage{bm}
\usepackage{mathrsfs}
\usepackage{times}

\allowdisplaybreaks

\begin{document}
\title{An exact solution for the Hawking effect in a dispersive fluid}

\author{T.\ G.\ Philbin}

\email{t.g.philbin@exeter.ac.uk}

\affiliation{Physics and Astronomy Department, University of Exeter,
Stocker Road, Exeter EX4 4QL, United Kingdom}

\begin{abstract}
We consider the wave equation for sound in a moving fluid with a fourth-order anomalous dispersion relation. The velocity of the fluid is a linear function of position, giving two points in the flow where the fluid velocity matches the group velocity of low-frequency waves. We find the exact solution for wave propagation in the flow. The scattering shows amplification of classical waves, leading to spontaneous emission when the waves are quantized. In the dispersionless limit the system corresponds to a 1+1-dimensional black-hole or white-hole binary and there is a thermal spectrum of Hawking radiation from each horizon. Dispersion changes the scattering coefficients so that the quantum emission is no longer thermal. The scattering coefficients were previously obtained by Busch and Parentani in a study of dispersive fields in de Sitter space [Phys.\ Rev.\ D {\bf 86}, 104033 (2012)]. Our results give further details of the wave propagation in this exactly solvable case, where our focus is on laboratory systems.
\end{abstract}

\maketitle

\section{Introduction}
The scattering of waves by material inhomogeneities occurs in many guises and is most familiar through the reflection of light at sharp boundaries. The basic phenomenon of scattering may seem intuitive, but new types of scattering become possible when the inhomogeneity consists of a position-dependent velocity of the medium. The flow of the medium can allow additional propagating modes that are absent in the non-moving case. Scattering into the additional modes may be accompanied by wave amplification in which energy is transferred  from the flow to the wave. When this amplification occurs in a region where the flow velocity exceeds the group velocity of the wave, the process is analogous to that underlying the Hawking effect at a black-hole horizon~\cite{unr81,bar05,rob12}. Inhomogeneous flow profiles and black-hole horizons can both act as amplifiers for waves. The quantum Hawking effect~\cite{haw74,bro95} of black holes is due to the coupling of the horizon amplifier to the wave's quantum ground state, the latter being ``amplified" into real quanta or quantum noise~\cite{Gardiner}. Both classical wave amplification and spontaneous quantum emission can occur in inhomogeneous flows and several theoretical proposals have now been investigated experimentally~\cite{gar00,sch02,rou08,phi08,bel10,lah10,wei11,ste14,ngu15,ste15,euv15a}. 

The Hawking effect in moving media differs most significantly from the astrophysical case in the essential role played by dispersion. The lack of short-wavelength dispersion for waves in space-time renders the astrophysical Hawking effect somewhat singular, as its derivation features infinite wavelength shifts at the horizon~\cite{Jacobson}. For waves in material media, dispersion necessarily limits the wavelength shifts and the Hawking effect has no unphysical features~\cite{unr95,cor96}. Dispersion has been shown to alter the spectrum of Hawking radiation from the original  thermal result for black holes~\cite{rob12,mac09,leo12,fin12,cou12,mic14,rob14,euv15}. In general the spectrum of quantum Hawking emission depends on the dispersion and the shape of the flow profile. Moreover a flow velocity that exceeds the group velocity of the wave is not necessary for measurable wave amplification to occur~\cite{wei11,euv15a,euv15}. Waves in moving media thus offer a rich theoretical and experimental arena where the quantitative dependence of the Hawking effect on dispersion and flow profile can be explored.

A less welcome effect of dispersion is to make wave scattering in the experimental systems very difficult to solve analytically. The most accessible experimental system to date uses surface waves on flowing water, where Hawking amplification of classical incident waves has been observed~\cite{wei11,euv15a}. The dispersion of water waves is complicated, with regions of normal dispersion ($d^2\omega/dk^2<0$ in the fluid frame) and anomalous dispersion ($d^2\omega/dk^2>0$ in the fluid frame) giving various types of horizon effects~\cite{sch02,rou10,pel15}. The system most studied theoretically is sound waves in a flowing Bose-Einstein condensate (BEC), and experiments with BECs have detected quantum Hawking emission of phonons~\cite{ste14,ste15}. A fourth-order dispersion relation is widely used to model the BEC system~\cite{bar05}, and a polynomial dispersion relation can also be used in a simple model for water waves. The resulting wave equations have not been exactly solved for an inhomogeneous flow and theoretical predictions of the Hawking effect in these systems are based on approximate analytical techniques and numerical simulations~\cite{unr95,cor96,bro95a,cor98,him00,sai00,unr05,mac09,mac09b,leo12,rob14,fin12,cou12,mic14,euv15,rob16,cou16,mic16}. 

Here we give the exact solution for wave scattering in a flow whose velocity changes linearly with position, where we use a simple BEC model featuring a fourth-order anomalous dispersion relation~\cite{bar05}. The one-dimensional linear flow profile has regions of positive and negative flow velocity, so there are two horizons, one for left-moving and one for right-moving waves. The flow is thus analogous to a black-hole or white-hole binary (depending on the sign of the velocity gradient). A linear flow profile is often assumed in the neighbourhood of a single horizon, and approximate treatments of this case using the same techniques employed here were given in~\cite{unr05,cou12}. But the exact treatment of the linear profile necessarily involves two horizons. On the other hand, the wave scattering does not depend on the profile being linear at large distances, so that the scattering coefficients obtained here are also valid for profiles that slowly change from linear to flat far from the horizons.

The linear flow profile in the dispersionless case gives an effective space-time metric for waves that corresponds to a patch of de Sitter space, as discussed in ref.~\cite{bus12}. In that work, Busch and Parentani studied dispersive fields in de Sitter space and in a cosmological context considered the dispersive wave equation used here. These authors derived the scattering amplitudes (\ref{S12})--(\ref{S42}) below by a somewhat different method. Here we provide more details on the exact wave solution and its asymptotics. Also, our focus here is on laboratory systems rather than cosmology.

The linear flow profile is not the example one would most like to solve exactly because it does not model current experiments, but nevertheless it gives some interesting lessons. The results highlight the importance of \emph{reflection}, the familiar conversion of right-moving waves to left-moving and vice versa, where left/right motion here refers to the velocity relative to the fluid. Scattering due to reflection in inhomogeneous flows has until recently been of secondary interest in studies of Hawking amplification, but reflection occurs in the experimental systems and its magnitude affects the spectrum of quantum Hawking emission~\cite{euv15}. For theoretical purposes, dispersion can be implemented in a manner that does not give reflection, so that the left- and right-moving waves do not mix~\cite{sch08}, but this possibility is not realised in the experimental systems.

For wave equations in spatially inhomogeneous media, the exact solution for a linearly changing profile is the basis for the WKB approximation for arbitrary profiles~\cite{heading}. In optics and quantum mechanics the solutions in question are the two Airy functions, valid for a linearly changing permittivity or potential. For the wave equation in a linearly inhomogeneous flow we will obtain four solutions, rather than two, and importantly  they depend on the dispersion. Our solutions do not therefore have the same universal significance for inhomogeneous flows as the Airy functions have in optics and quantum mechanics, rather they are specific to the dispersion of the BEC model.

The model equation used together with a general description of the propagating modes is given in section~\ref{sec:wave}. In section~\ref{sec:ksol} the wave equation is Fourier transformed and solved exactly in $k$-space. Section~\ref{sec:4sols} presents the lengthiest part of the analysis, in which four independent solutions of the wave equation are defined and their meaning in terms of mode scattering is elucidated. In section~\ref{sec:scat} solutions are constructed that represent the scattering of single incident modes and the exact scattering coefficients are calculated. The final results are contained in (\ref{S12})--(\ref{S42}) and (\ref{S2S3a})--(\ref{S2S3b}), and are visually represented in Figs.~\ref{fig:2in} and~\ref{fig:3in}.

\section{Wave equation}   \label{sec:wave}
An approximate (1+1)-dimensional wave equation for sound in a flowing BEC takes the form~\cite{bar05} 
\begin{equation}  \label{wave}
\partial_t (\partial_t+v\partial_x )\psi+\partial_x (v\partial_t+v^2\partial_x )\psi-\left(\partial_x^2-\frac{1}{k_c^2}\partial_x^4\right)\psi=0.
\end{equation}
Here velocity is dimensionless, $v(x)$ is the (time-independent) flow velocity, $k_c$ quantifies a dispersive term, and $\psi(x,t)$ is the wave field; for the BEC $\psi(x,t)$ is the phase fluctuation in the field operator of the bose gas~\cite{bar05}. A more accurate equation for sound in a BEC can be derived~\cite{mac09b}, but (\ref{wave}) is often considered in studies of dispersion and the Hawking effect.  In the absence of the fourth-order derivative term there is no dispersion of the wave and  (\ref{wave}) is the equation of a scalar field in a curved (1+1)-dimensional space-time~\cite{unr81,bar05}. In the non-dispersive case ($k_c\to\infty$) the waves have speed $1$ relative to the fluid and points where $v(x)=\pm1$ are in strict analogy to event horizons for the waves~\cite{unr81}. 

We take (\ref{wave}) as our model wave equation in a dispersive fluid. The central problem is to solve (\ref{wave}) as a classical wave equation, as this determines the mode expansion for $\psi(x,t)$ as a classical or quantum field. In particular, the Hawking effect in this system is at root a classical scattering effect for waves satisfying (\ref{wave}). The monochromatic wave equation follows from the substitution $\psi(x,t)=e^{-i\omega t} \phi(x)$:
\begin{equation}  \label{mono}
 \left[\omega^2  +i\omega v'+2 v (i\omega-v')\partial_x   +(1-v^2) \partial_x^2 -\frac{1}{k_c^2}\partial_x^4   \right]\phi=0,
\end{equation}
where a prime denotes a derivative with respect to $x$. The dispersion relation is
\begin{equation} \label{disprel}
(\omega-vk)^2=k^2+\frac{k^4}{k_c^2},
\end{equation}
which can be solved exactly for four roots $k(\omega)$ that can all describe propagating modes in the fluid for $\omega>0$ (the expressions for $k(\omega)$ are too cumbersome to reproduce here). The frequency $\omega$ in the laboratory frame is conserved for monochromatic waves because of the time-independence of (\ref{mono}). The quantity $\omega-v(x)k$ is the frequency in a frame locally co-moving with the fluid, which is the relevant frame for characterising the material dispersion. We see that (\ref{disprel}) gives anomalous dispersion at all frequencies, so the group and phase speeds relative to the fluid exceed their non-dispersive values of $1$, except in the limit of zero $k$. As long as the dispersion is entirely anomalous or entirely normal for all frequencies, there will still be at most four propagating modes with $\omega>0$. If the dispersion is normal for some frequency ranges and anomalous for others, there are in general more than four $\omega>0$ propagating modes (water waves provide an example~\cite{rou10}). 

\begin{figure}[!htbp]
\includegraphics[width=\linewidth]{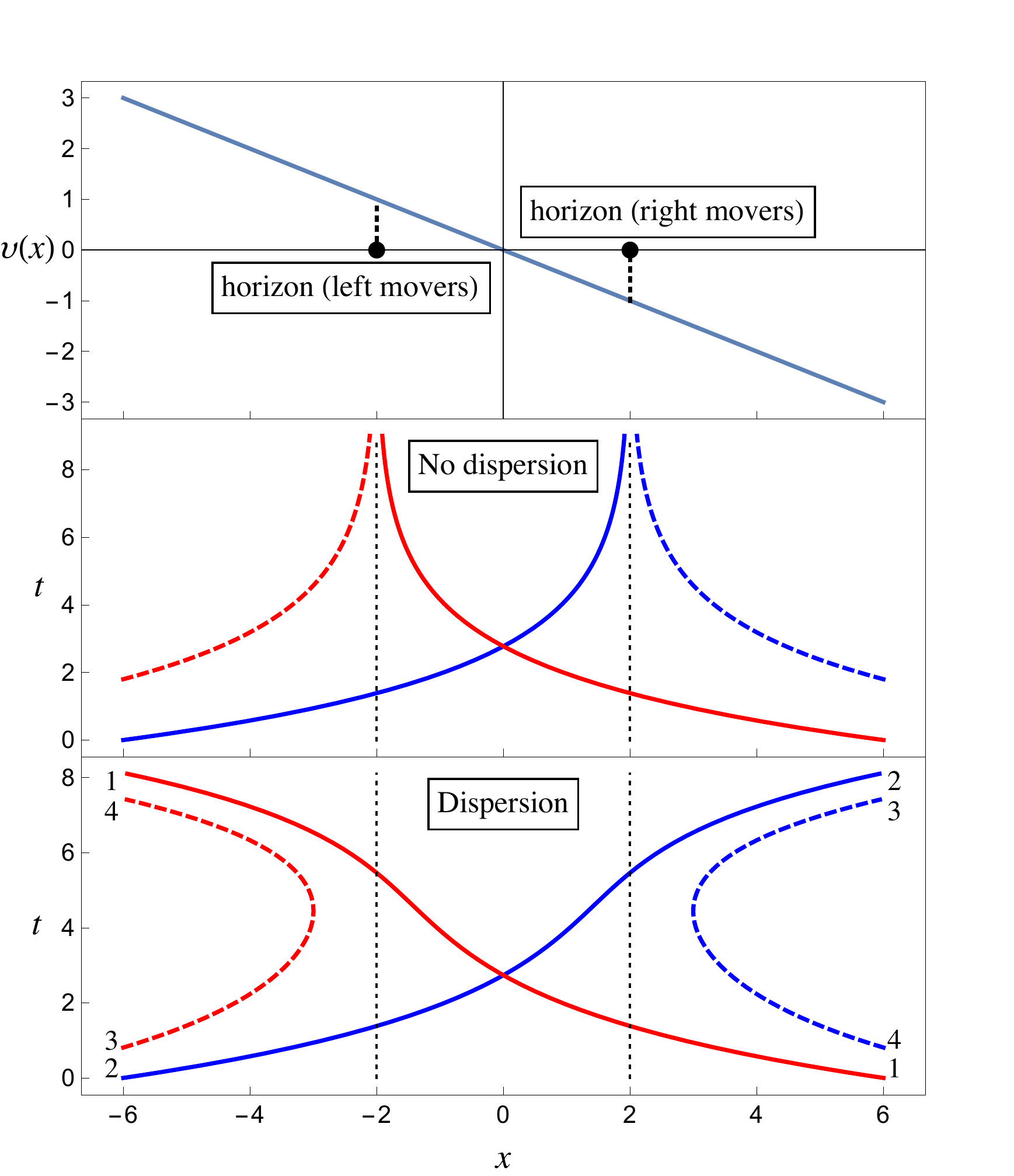}   
\caption{Top: the linear velocity profile (\ref{flow}) for $\alpha=1/2$. At $x=\pm1/\alpha$ the flow speed matches the speed relative to the fluid of low-$k$ waves. Middle: ray plots for waves in the flow with $\omega=1$, and with dispersion removed ($k_c\to\infty$). Red (blue) rays are for waves moving left (right) relative to the fluid. Solid (dotted) rays have positive (negative) frequency $\omega-v(x)k$ in a frame co-moving with the fluid. This corresponds to a white-hole binary. Bottom: the ray plots when dispersion is included ($k_c=5$). The mode labels 1, 2, 3 and 4 refer to the four solutions $k(\omega)$ of the dispersion relation (\ref{disprel}); these solutions are shown graphically in Fig.~\ref{fig:disp}.
} \label{fig:rays}
\end{figure}

\begin{figure}[!htbp]
\includegraphics[width=\linewidth]{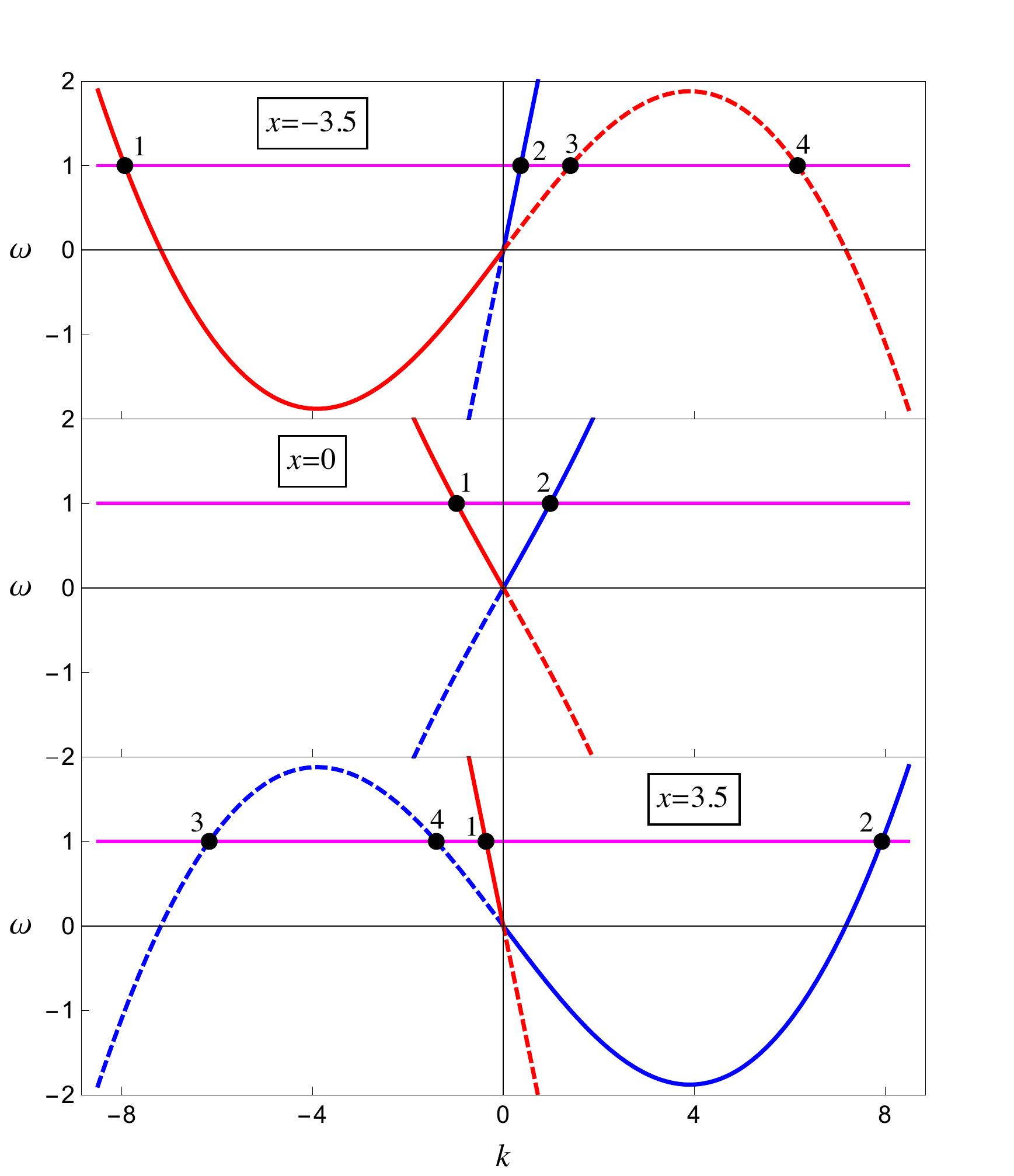}   
\caption{Plots of $\omega$ versus $k$ for the dispersion relation (\ref{disprel}) in the linear profile (\ref{flow}), for three values of $x$ (with $\alpha=1/2$, $k_c=5$). The horizontal magenta line is $\omega=1$ and its intersection with the dispersion plots gives the propagating (real $k$) modes at this frequency; these mode solutions are labeled 1 to 4. Modes from an intersection with a red (blue) dispersion curve travel left (right) relative to the fluid. In addition, if the intersection is with a solid (dotted) dispersion curve, the mode has positive (negative) co-moving frequency $\omega-v(x)k$. In the region between the horizons, such as in the middle plot for $x=0$, there are two propagating modes. Beyond the horizons ($x\ll 0$ and $x\gg0$) modes 3 and 4 can propagate.
} \label{fig:disp}
\end{figure}

We consider the linear flow profile
\begin{equation} \label{flow}
v(x)=-\alpha x, \qquad \alpha>0
\end{equation}
(see Fig.~\ref{fig:rays}). This flow exceeds the small-$k$ (non-dispersive) speed of the waves relative to the fluid at $x=\pm1/\alpha$. In the non-dispersive limit $k_c\to\infty$ the flow provides a sharp horizon for left-going waves at  $x=-1/\alpha$, and for right-going waves at $x=1/\alpha$. This is shown in Fig.~\ref{fig:rays}, where the rays for $\omega>0$ waves are plotted in the non-dispersive case. The two rays for left movers (relative to the fluid) are shown in red and are funnelled to the horizon at $x=-1/\alpha$, undergoing an infinite blue-shifting ($k\to\infty$) in the process. The solid (dotted) ray has a positive (negative) frequency $\omega-v(x)k$ in the co-moving frame. Similarly the two right-movers (relative to the fluid), one of which has a negative co-moving frequency, are funnelled to the horizon at $x=1/\alpha$ and infinitely blue-shifted. This flow profile corresponds to a white-hole binary, i.e.\ two 1+1-dimensional white holes facing each other. By reversing the direction of time (or reading the ray plot from top to bottom) a black-hole binary is obtained. The anomalous dispersion in the wave equation (\ref{mono}) limits the blue-shifting at the horizons and redirects the blue-shifted left- and right-movers into the respective white holes, as also shown in Fig.~\ref{fig:rays}. This behaviour can be understood by looking at the dispersion relation at different point in the flow, shown in Fig.~\ref{fig:disp}. As noted above, there are four propagating modes $k(\omega)$ in the flow, labeled 1 to 4 in the bottom ray plot in Fig.~\ref{fig:rays} and in the graphical solutions for these modes shown in Fig.~\ref{fig:disp}. Modes 1 and 2 are the usual left- and right-moving $\omega>0$ waves that would be present in a nonmoving fluid ($v(x)=0$); these modes have positive frequency in a frame co-moving with the fluid. The motion of the fluid has the remarkable effect of making modes 3 and 4 into propagating $\omega>0$ waves, despite their having a negative co-moving frequency $\omega-v(x)k$. Modes 3 and 4 have complex wave-vectors in the region between the horizons, with mode 3 exponentially increasing with $x$ in this region and mode 4 exponentially decreasing with $x$. It is not visually apparent from Figs.~\ref{fig:rays} and~\ref{fig:disp} how to follow modes 3 and 4 through the region where their wave-vectors are complex and  identify which is mode 3 and which is mode 4 when they again become propagating. The  identification follows from the four exact roots of the quartic dispersion relation (\ref{disprel}) and a choice of branch cuts in these roots. Modes 3 and 4 coincide at two points in the flow, as is clear from the bottom ray plot in Fig.~\ref{fig:rays}; these modes are thus converted into each other in a continuous manner by the flow. Rays move at the group velocity $d\omega/dk$, and Fig.~\ref{fig:disp} explains the reversal of group velocity as modes 3 and 4 are converted into each other. Note also that when mode 4 propagates in the left-hand region of the flow its phase ($\omega/k$) and group ($d\omega/dk$) velocities are in opposite directions, and similarly for mode 3 in the right-hand region. 

Mixing of modes 3 and 4 occurs at the level of ray tracing and is therefore visible in the bottom ray plot in Fig.~\ref{fig:rays}. In addition, there is also wave scattering of each of the four modes into the other three modes because of a breakdown of geometrical optics near the horizons. Our goal is to calculate the exact scattering coefficients. The scattering of modes 1 and 2, with positive co-moving frequency, into modes 3 and 4, which have negative co-moving frequency, is accompanied by wave amplification and is the underlying mechanism of the Hawking effect~\cite{unr95}. Modes 3 and 4 scatter into modes 1 and 2 with a similar amplification effect. The wave amplification can be partly understood in terms of the conserved quantities associated with the wave equation (\ref{wave}). As detailed in Appendix~\ref{app:conserved}, there is a conserved norm, associated with $U(1)$ symmetry of the action giving (\ref{wave}), and a conserved pseudo-energy associated with its time-translation invariance. The sign of the norm and pseudo-energy of each mode is given by the sign of its co-moving frequency (see Appendix~\ref{app:asymptotic}). When a mode with positive co-moving frequency scatters into a mode with negative  co-moving frequency, the total norm can only be conserved if the original mode \emph{increases} its norm. There is thus an amplification of the wave excitation in the system. This occurs with conservation of pseudo-energy, but in the real physical system the  energy of the wave has increased while energy is removed from the fluid motion (this corresponds to evaporation of black holes~\cite{haw74,bro95}). A complete treatment of the physics involved would require a full account of back-reaction on the fluid together with the resulting energy transfer from the flow to the wave, but in (\ref{wave}) the fluid appears simply as an external field $v(x)$. In actual experiments the amplification effect is so small that ignoring back-reaction in theoretical predictions is justified.

In order for the notion of scattering coefficients to make sense it is necessary for the waves in the asymptotic regions $|x|\to\infty$ to reduce to non-interacting superpositions of waves associated with the four roots of the dispersion relation and their corresponding rays. In discussing the ``modes" labeled 1 to 4 above, we have implicitly assumed the validity of such a picture. If the flow has constant velocity in the asymptotic regions $|x|\to\infty$, it is clear the waves will asymptotically become superpositions of non-interacting plane waves given by the dispersion relation. For the linear profile (\ref{flow}), however, it is not immediately obvious that the breakdown of geometrical optics is confined to the horizon regions, since the profile has the same rate of change at all points $x$ and there are rays whose wavelengths grow longer as $|x|$ increases (mode 2 on the left and mode 1 on the right). Nevertheless, waves in the linear profile in the regions $|x|\to\infty$ do reduce to non-interacting superpositions whose components are associated with the roots of the dispersion relation and their corresponding rays. This is shown in Appendix~\ref{app:asymptotic}, where it is also found that the low-$k$ asymptotic modes do not have the same dependence on $v(x)$ as they do for asymptotically constant flows. (This last fact contrasts with the optical and quantum mechanical case, where the WKB solutions for the linear profile have exactly the same dependence on the profile function as for asymptotically constant profiles~\cite{heading}.) The flow regions on the far left and far right are thus places where we can legitimately speak of input and output waves and compute well-defined scattering coefficients. 

\section{Solution of the wave equation in $k$-space}   \label{sec:ksol}
We express the monochromatic wave $\phi(x)$ by a Fourier representation
\begin{equation}  \label{four}
\phi(x)=\int_C dk \, \tilde{\varphi}(k) e^{ik x},
\end{equation}
where we allow any contour $C$ in the complex $k$-plane such that the integral converges and $\tilde{\varphi}(k) e^{ik x}$ vanishes at the endpoints (in practice our contours will run to infinity in the complex plane). The wave equation (\ref{mono}) for the linear profile (\ref{flow}) then gives
\begin{align}
\left[k^2\left(1+\frac{k^2}{k_c^2}\right)-i\alpha\omega-\omega^2  \right] \tilde{\varphi}(k)  & +  2k\alpha(\alpha-i\omega)\tilde{\varphi}'(k)    \nonumber \\
& +k^2\alpha^2\tilde{\varphi}''(k)=0,
\end{align}
where a prime denotes a derivative. Extracting a factor through $\tilde{\varphi}(k)=k^{-1+i\omega/\alpha}\exp[-ik^2/(2\alpha k_c)]f(k)$ and rescaling the variable $k$ leads to a differential equation of the well-studied form
\begin{equation}
f''(z)-2 z f'(z)+af(z)=0.
\end{equation}
We thereby obtain the general solution
\begin{align}
\tilde{\varphi}(k)=k^{-1+\frac{i\omega}{\alpha}} &  \exp\left(- \frac{ik^2}{2\alpha k_c} \right)  \left[ c_1   H_{-\frac{1}{2}-\frac{ik_c}{2\alpha}}\left(\sqrt{\frac{i}{\alpha k_c}} \, k \right)   \right.   \nonumber \\
& \qquad \left. + c_2\,_1\!F_1\left(\frac{1}{4}+\frac{ik_c}{4\alpha};\frac{1}{2};\frac{ik^2}{\alpha k_c}  \right)  \right],    \label{ksol}
\end{align}
where $H_a(z)$ is the Hermite function (which is a polynomial for non-negative integer $a$), $_1\!F_1(a;b;z)$ is the Kummer confluent hypergeometric function, and $\{c_1,c_2\}$ are arbitrary constants.

Through (\ref{four}) and (\ref{ksol}) we now have the general solution for the wave $\phi(x)$, expressed as an  integral representation. Although there are only two linearly independent solutions in (\ref{ksol}) for $ \tilde{\varphi}(k)$, there are four independent solutions for $\phi(x)$ because of the freedom to choose a branch cut in the integrand in  (\ref{four}). We choose two independent solutions for $ \tilde{\varphi}(k)$ by taking just the Hermite term, or just the hypergeometric term, in (\ref{ksol}). For each of these  $ \tilde{\varphi}(k)$ we will obtain two independent solutions for $\phi(x)$ by the choice of the contour and branch cut in (\ref{four}).

\section{Four independent solutions of the wave equation}   \label{sec:4sols}
Both the Hermite  function $H_a(z)$ and the confluent hypergeometric function $_1\!F_1(a;b;z)$  are entire functions of $z$~\cite{dlmf}. The $k$-space solution (\ref{ksol}) therefore has one singularity and one branch cut in the complex $k$-plane due to the factor $k^{-1+\frac{i\omega}{\alpha}}$. In the contour integral (\ref{four}), which gives us the exact solutions to the wave equation, we must avoid the singularity at $k=0$ and negotiate the branch cut. We will choose the branch cut to run either along the positive or the negative imaginary $k$-axis. Then we choose the contour in (\ref{four}) to run from $k=-\infty$ to $k=\infty$, as in the usual Fourier transform, but avoiding $k=0$ by running below $k=0$ (branch cut along positive imaginary axis) or above $k=0$ (branch cut along negative imaginary axis). These two choices of branch cut and contour will give two independent solutions for $\phi(x)$.

\subsection{Two solutions with $H_a(z)$ in their integral representations}
\begin{figure}[!htbp]
\includegraphics[width=\linewidth]{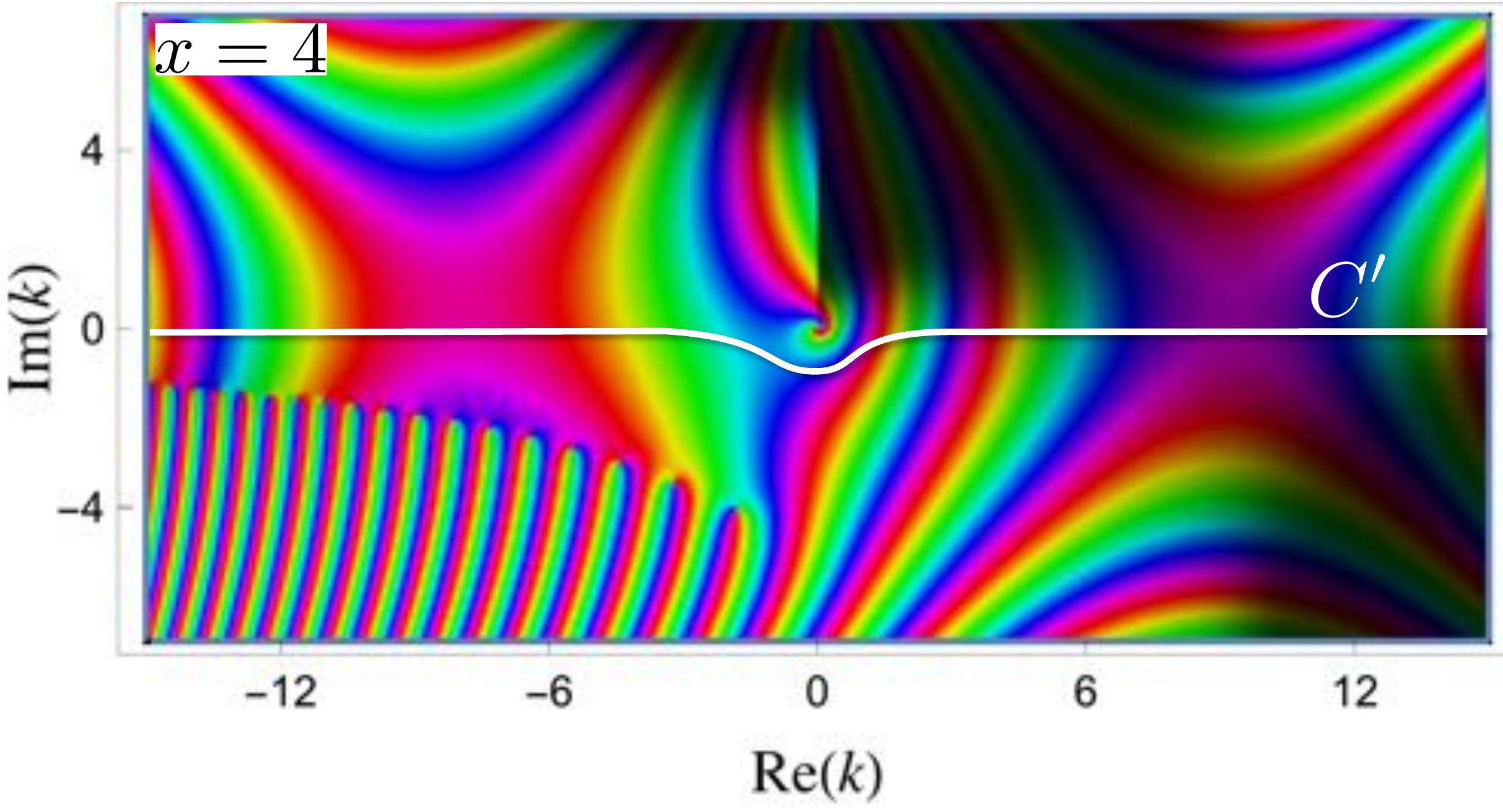}   
\includegraphics[width=\linewidth]{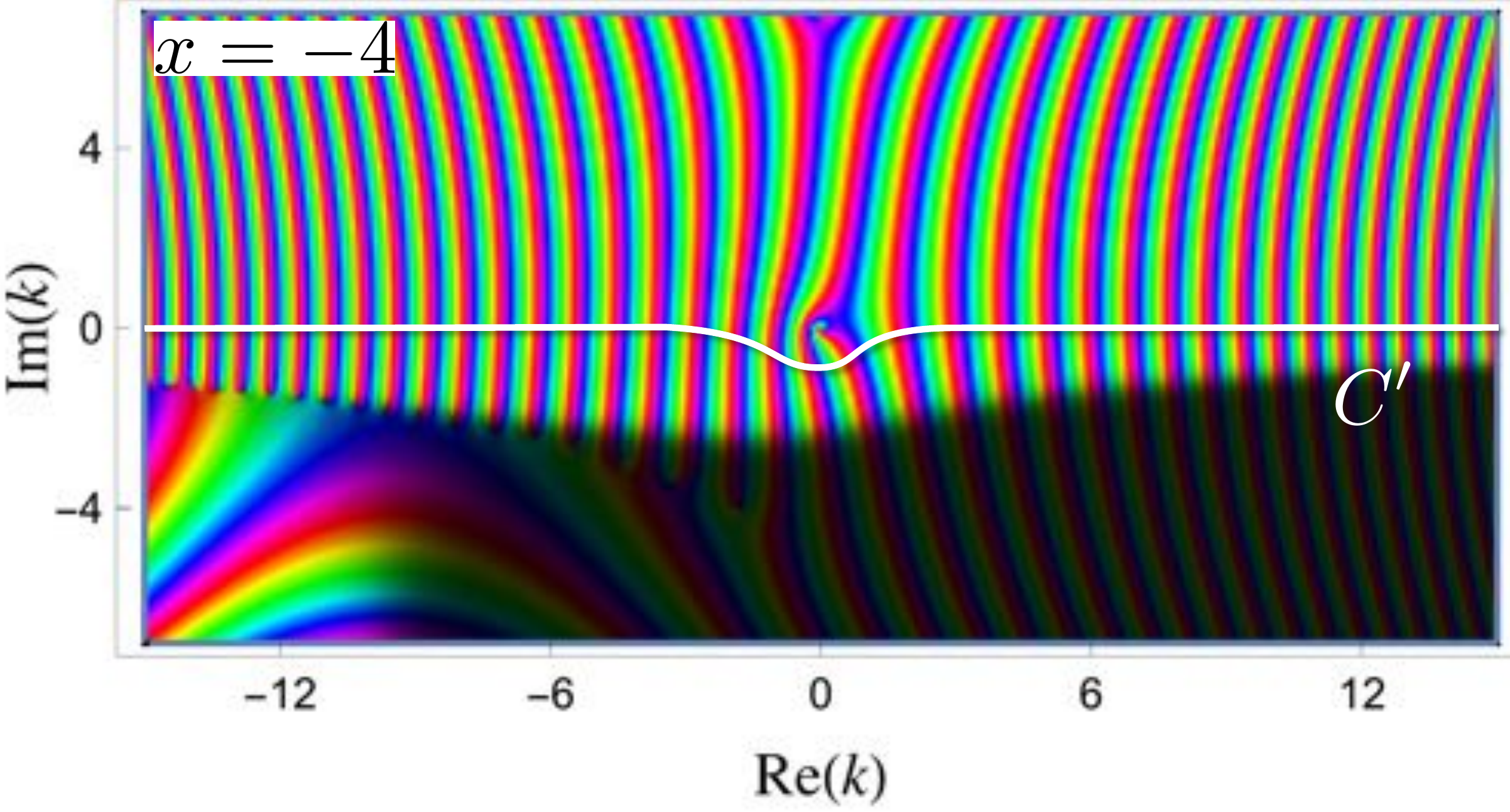}   
\caption{The integrand in (\ref{four}) plotted in the complex $k$-plane, with $c_1=1$ and $c_2=0$ in (\ref{ksol}). Colour indicates the phase while brightness indicates the absolute value (brighter means bigger). The branch cut is placed along the positive imaginary axis. Parameter values are $\omega=1$, $\alpha=1/2$ and $k_c=5$. The top plot is for $x=4$, the bottom for $x=-4$. We take $C'$ as the contour in (\ref{four}) and thereby define an exact solution of (\ref{mono}).
} \label{fig:Hermpos}
\end{figure}

\begin{figure}[!htbp]
\includegraphics[width=\linewidth]{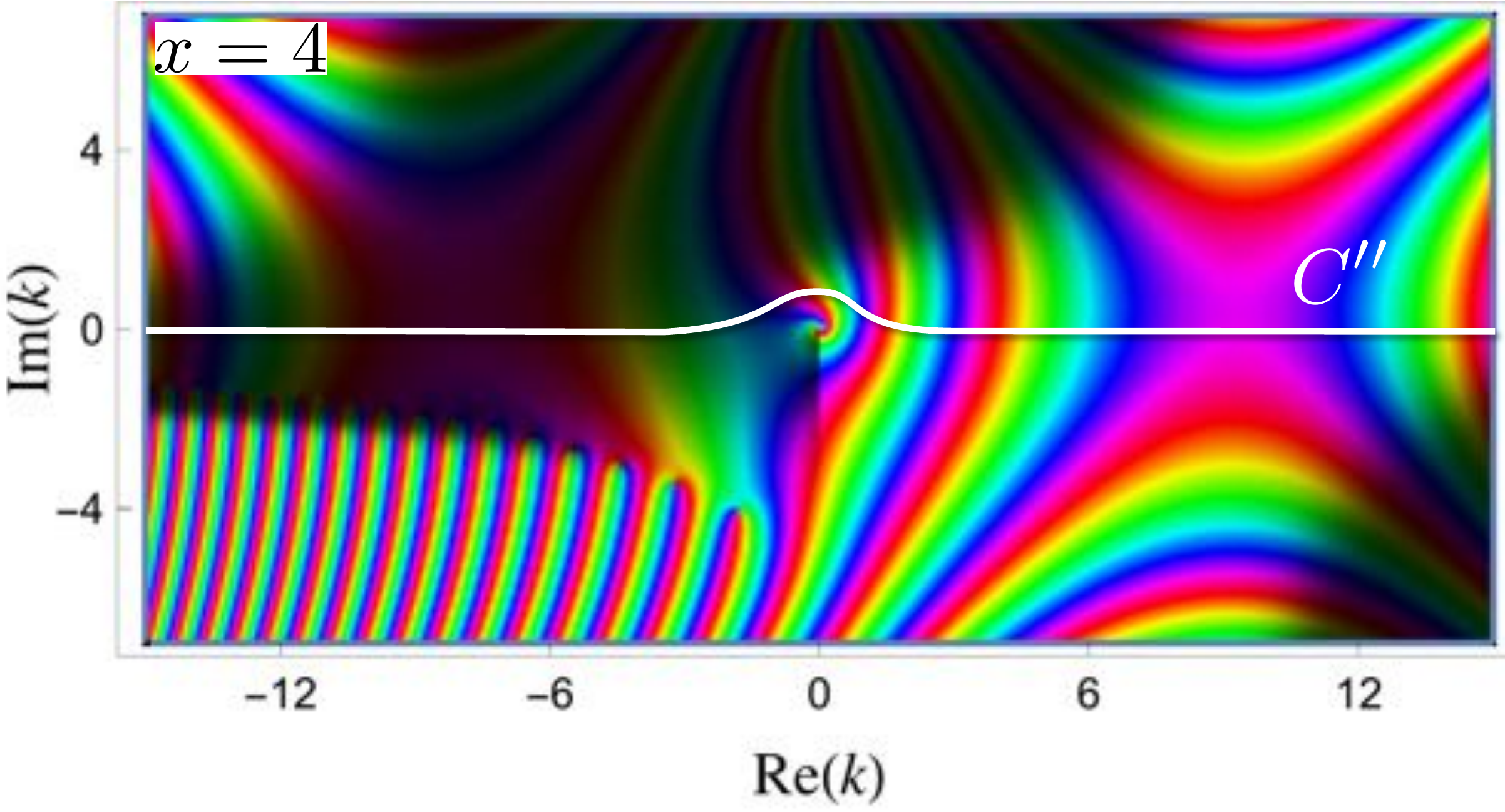}   
\includegraphics[width=\linewidth]{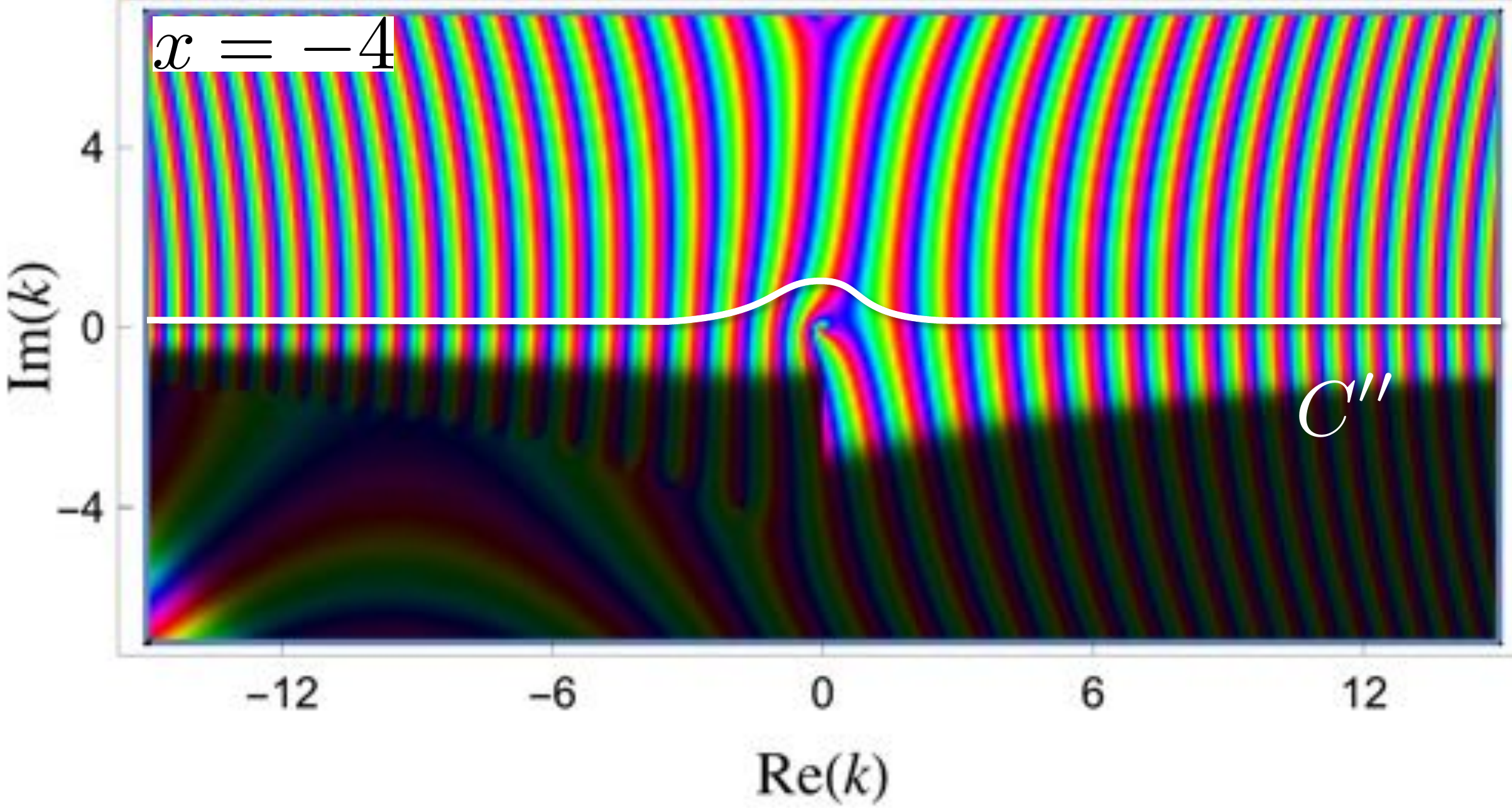}   
\caption{The same as Fig.~\ref{fig:Hermpos}, but with the branch cut along the negative imaginary axis and a different contour $C''$. This defines another independent solution of (\ref{mono}).
} \label{fig:Hermneg}
\end{figure}

With $c_1=1$ and $c_2=0$ in (\ref{ksol}), the integrand in (\ref{four}) contains the Hermite  function $H_a(z)$. This integrand is plotted in the complex $k$-plane in Figs.~\ref{fig:Hermpos} and~\ref{fig:Hermneg}, with the branch cut running along the positive or negative imaginary axis, respectively. Also shown in each figure is a contour that  that defines a solution (\ref{four}). The contours are as described above and labeled $C'$ (branch cut along positive imaginary axis) or $C''$ (branch cut along negative imaginary axis).

Our main goal is to quantify the wave scattering in the flow and for this we need the asymptotic expansions of the exact solutions (\ref{four}) in the large $|x|$ regions. Every exact solution must reduce as $|x|\to\infty$ to a superposition of the asymptotic wave components derived in Appendix~\ref{app:asymptotic}. These asymptotic waves do not interact and correspond to the four ray solutions of Sec.~\ref{sec:wave}, propagating in the large $|x|$ regions.

If the contour in (\ref{four}) ran along the real $k$-axis we would have a Fourier transform.  For large $|x|$, the rapid oscillation in $k$ of the $e^{ikx}$ factor tends to make the Fourier transform zero as $|x|\to\infty$, provided the function being transformed is well behaved. The Riemann-Lebesgue lemma gives the technical requirements~\cite{titchmarsch}. In our case there is a singularity at $k=0$ and we do not integrate along the real $k$-axis, so we need not expect the integral (\ref{four}) to vanish as $|x|\to\infty$. Indeed, we know that the low-$k$ asymptotic wave components do not vanish as  $|x|\to\infty$ (see Appendix~\ref{app:asymptotic}). Hence the integral (\ref{four}) will be nonzero as $x\to-\infty$ ($x\to\infty$) if and only if the solution $\phi(x)$ contains any of the low-$k$ asymptotic wave components in the far left (far right) regions. The high-$k$ asymptotic wave components, on the other hand, vanish as $|x|\to\infty$ (see Appendix~\ref{app:asymptotic}) and we must also extract these components from the solution (\ref{four}). We thus require not simply the leading-order part of the solution (\ref{four}) for large $|x|$, but rather the leading-order contributions to all asymptotic wave components that are present.

To find the asymptotic expansions for $x\to\pm\infty$ of the integral (\ref{four}), we must consider the behaviour of the integrand along the contour. The contours $C'$ and $C''$ in Figs.~\ref{fig:Hermpos} and~\ref{fig:Hermneg} run mostly along the real $k$-axis. For large $|x|$, the rapid oscillation of $e^{ikx}$ gives net cancellation for the parts of the contours on the real axis and so their contribution to the integral will vanish as $|x|\to\infty$. The leading-order term in these contributions is determined by whether or not there are points of stationary phase in the integrand that lie close to the contour. One can see two patches in the top plots in Figs.~\ref{fig:Hermpos} and~\ref{fig:Hermneg} where the phase has an extremum in the complex plane, while in the lower plots the integrand appears to be the sum of two parts, one with an extremum of the phase that is situated to the left and another part that has no extrema. Moreover all these extrema move out along the real $k$-axis as $|x|$ increases, so to investigate them analytically we must consider the form of the integrand for large $|k|$. We will find below that there are terms in the integrand with points of stationary phase that are precisely the points we have just identified visually. Moreover each of these stationary-phase points is the wave-vector of one of the high-$k$ asymptotic waves and the resulting contributions to the integral will give all the high-$k$ asymptotic wave components. There remains the parts of the contours  $C'$ and $C''$ that do not lie on the real $k$-axis. For $x>0$ the factor $e^{ikx}$ decreases exponentially along the positive imaginary $k$-axis, whereas for $x<0$ it decreases along the negative imaginary axis. All other factors in the integrand are dominated by this behaviour, as is visible in Figs.~\ref{fig:Hermpos} and~\ref{fig:Hermneg}. We can deform the parts of the contours $C'$ and $C''$ that leave the real axis so that as much as possible they lie along the half of the imaginary axis where the integrand is exponentially small. If there is no branch cut along the relevant half of the imaginary axis (as in the bottom plot of Fig.~\ref{fig:Hermpos} and the top plot of Fig.~\ref{fig:Hermneg}), then the contribution of this part of the contour is arbitrarily small. If however the branch cut lies along the half of the imaginary axis where the integrand is exponentially decreasing (as in the top plot of Fig.~\ref{fig:Hermpos} and the bottom plot of Fig.~\ref{fig:Hermneg}), then the contour gets wrapped around $k=0$ and runs along both sides of the branch cut. The latter situation will give a contribution to the integral that does not vanish as $|x|\to\infty$ and corresponds to low-$k$ asymptotic wave components present in the solution. This outlines how the asymptotic wave components are encoded in the exact integral representation (\ref{four}). A similar identification of wave components occurs in~\cite{unr05}, where the techniques used here were applied to an approximate treatment of the wave equation in an arbitrary flow profile.  

To identify the points of stationary phase discussed above, we require the asymptotic expansions for $k\gg 0$ and $k\ll 0$ of the Hermite function that appears in (\ref{ksol}). Asymptotic expansions of the Hermite function for large variable are given in~\cite{dlmf}; in our case the variable is $\sqrt{i/(\alpha k_c)}k$, giving the following leading-order expansions:
\begin{widetext}
\begin{align}
H_{-\frac{1}{2}-\frac{ik_c}{2\alpha}}\left(\sqrt{\frac{i}{\alpha k_c}}k \right)   \sim  & \, 2^{-\frac{1}{4}-\frac{ik_c}{4\alpha}}  \left( \sqrt{\frac{2i}{\alpha k_c}} \, k \right)^{ -\frac{1}{2}-\frac{ik_c}{2\alpha} }, \quad -\pi < \text{arg}(k) <\frac{\pi}{2},    \label{Hposk}   \\
H_{-\frac{1}{2}-\frac{ik_c}{2\alpha}}\left(\sqrt{\frac{i}{\alpha k_c}}k \right)   \sim  & \,  2^{-\frac{1}{4}-\frac{ik_c}{4\alpha}}  \left[  \left( \sqrt{\frac{2i}{\alpha k_c}} \, k \right)^{ -\frac{1}{2}-\frac{ik_c}{2\alpha} }  \right.
 \nonumber   \\
&   \left. - \frac{i \sqrt{2\pi}}{\Gamma\left( \frac{1}{2}+\frac{ik_c}{2\alpha}  \right)}   \left( \sqrt{\frac{2i}{\alpha k_c}} \, k \right)^{ -\frac{1}{2}+\frac{ik_c}{2\alpha} }   \exp\left( \frac{i k^2}{\alpha k_c} - \frac{\pi k_c}{2\alpha} \right)   \right]  ,    \quad -2\pi < \text{arg}(k) < -\frac{\pi}{2}.      \label{Hnegk}
\end{align}
\end{widetext}
Here the branch cuts in the asymptotic expressions are rotated out of the range of $\arg(k)$ for which they are valid. This amounts to analytically continuing $\ln z$ from its principal branch $-\pi<\arg(z)\leq \pi$. 

The points of stationary phase in the integrand in (\ref{four}) can now be identified (where here we take just the Hermite-function term in (\ref{ksol})). In the region where (\ref{Hposk}) is valid, which includes the positive real $k$-axis, the integrand has an exponential factor containing $k$ given by $\exp\left[ikx- ik^2/(2\alpha k_c) \right]$. This factor has a point of stationary phase at
\begin{equation} \label{stat1}
k=\alpha k_c x, \qquad x>>0,
\end{equation}
where the condition on $x$ occurs because (\ref{Hposk}) is not valid for negative real $k$. This point of stationary phase is visible on the right of the top plots in Figs.~\ref{fig:Hermpos} and~\ref{fig:Hermneg}. In the region where (\ref{Hnegk}) is valid, which includes the negative real $k$-axis, we have two terms in the asymptotic expansion of the Hermite function, the second of which has an exponential involving $k$. We must therefore treat the two terms in (\ref{Hnegk}) separately as they have different phase factors. The first term in (\ref{Hnegk}) gives a contribution to (\ref{four}) that has a phase factor $\exp\left[ikx- ik^2/(2\alpha k_c) \right]$, giving a point of stationary phase at
\begin{equation}  \label{stat2}
k=\alpha k_c x, \qquad x<<0,
\end{equation}
where now the condition is for $x$ to be negative because (\ref{Hnegk}) is not valid on the positive real $k$-axis. The second term in (\ref{Hnegk}) leads to a contribution to (\ref{four}) with a phase factor $\exp\left[ikx+ ik^2/(2\alpha k_c) \right]$, with stationary phase at
\begin{equation}  \label{stat3}
k= - \alpha k_c x, \qquad x>>0.
\end{equation}
The significance of the point (\ref{stat2}) is partly visible on the left of the lower plots in Figs.~\ref{fig:Hermpos} and~\ref{fig:Hermneg} but the integrand also has the contribution of the second term in (\ref{Hnegk}), for which there are no points of stationary phase for negative $x$. The point (\ref{stat3}) is similarly a stationary-phase point for only one term in the integrand but its significance is nevertheless apparent on the left of the top plots in Figs.~\ref{fig:Hermpos} and~\ref{fig:Hermneg}.

The points (\ref{stat1})--(\ref{stat3}) lie on the contours $C'$ and $C''$. For large $|x|$, we use the method of steepest descent to compute the the leading-order contribution to the integral of the terms which have a stationary phase at these points. Note that the points (\ref{stat1})--(\ref{stat3}) are the local wave-vectors of asymptotic wave components (see Appendix~\ref{app:asymptotic}) whose amplitudes fall off as $1/|x|^{3/2}$. We will find that the stationary-phase contributions give exactly these asymptotic wave components in a particular superposition. Contributions to this part of the integral that are not stationary-phase contributions fall off faster in $|x|$ than the stationary-phase contributions and therefore they are not needed to compute the asymptotic wave components. In the method of steepest descent~\cite{ablowitz} we expand the phase around the stationary point to second order and find the directions in the complex plane in which the quadratic term in the expansion is real and negative. We deform the contour to run along this line of steepest descent. Other factors in the integrand are evaluated at the stationary point and the resulting Gaussian integral along the steepest-descent line gives the leading-order contribution to the integral. In our case the phase factors are $\exp\left[ikx\mp ik^2/(2\alpha k_c) \right]$, which are simple to work with and give steepest-descent lines running through the stationary-phase points at angles of $\pm45^\circ$. When evaluating the other factors in the integrand at the stationary-phase point we must remember to choose the branch cut in the $k^{-1 + \frac{i\omega}{\alpha}}$ factor from (\ref{ksol}) to lie on the positive or negative imaginary axis, depending on which of the two solutions we are evaluating. This will complete the leading-order contributions to the solution from the parts of the contours $C'$ and $C''$ that lie on the real $k$-axis. There remains the portions of the contours that run around the singularity at $k=0$.

\subsubsection{First solution: branch cut along positive imaginary $k$-axis}
The branch cut in the integrand is first chosen to lie along the positive imaginary $k$-axis (Fig.~\ref{fig:Hermpos}) and the solution is defined by the contour $C'$. We denote this solution by $\phi^{(a)}(x)$. There are two expansions of  $\phi^{(a)}(x)$ in terms of the asymptotic waves, one for $x\ll 0$ and one for $x\gg 0$. 

Taking first $x\ll 0$, we see from (\ref{stat1})--(\ref{stat3}) that there is just one stationary-phase contribution, from the point (\ref{stat2}). This point is the wave-vector of mode 1 in the far-left region, whose normalized asymptotic form is $\phi_1^-(x)$ given by (\ref{-1}). The part of the contour $C'$ that leaves the real $k$-axis can be pushed down along the negative imaginary $k$-axis (see bottom plot in Fig.~\ref{fig:Hermpos}) where the integrand is exponentially small and gives no contribution. The only asymptotic wave component for $x\ll 0$ is thus mode 1 with a constant complex amplitude that we obtain from the method of steepest descent. The expansion of the solution is
\begin{gather}
\phi^{(a)}(x) \stackrel{x\ll 0}{\sim} a_1^- \phi_1^-(x) ,    \label{H1left}  \\
a_1^-=(-1)^\frac{1}{8}\sqrt{\pi}2^{\frac{1}{2}-\frac{ik_c}{2\alpha}}e^{\frac{\pi\omega}{\alpha}-\frac{3\pi k_c}{8\alpha}}\alpha (\alpha k_c)^{\frac{i\omega}{\alpha}-\frac{ik_c}{4\alpha}-\frac{1}{4}}.  \label{a1-}
\end{gather}

\begin{figure}[!htbp]
\includegraphics[width=7cm]{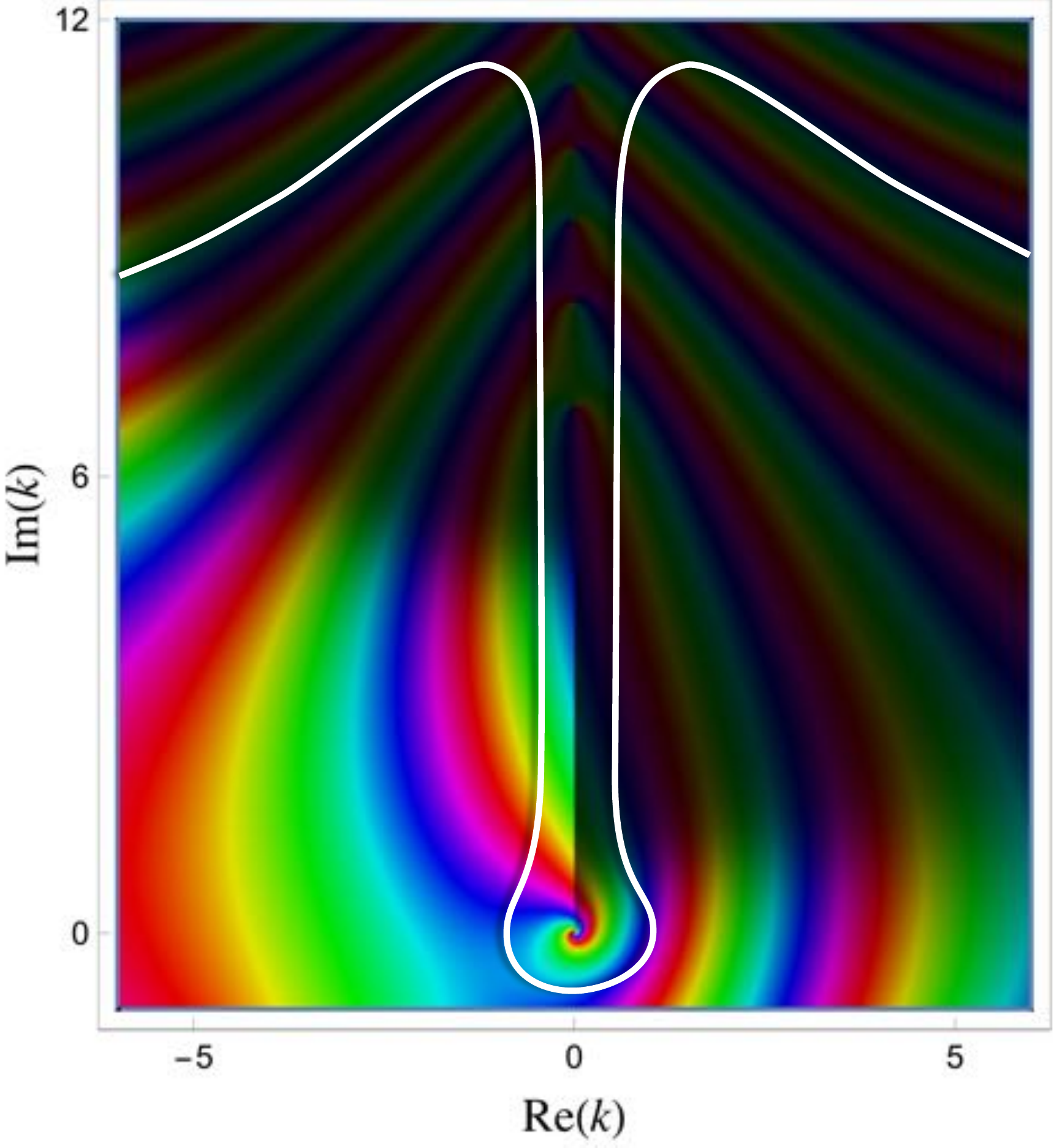}   
\caption{A portion of the upper plot in Fig.~\ref{fig:Hermpos}, with the contour $C'$ deformed to wrap around the singularity at $k=0$. The integrand decays exponentially along the positive imaginary axis. For the portion of the contour shown here, only the part running along both sides of the branch cut and around the singularity contributes to the integral for large $x$. 
} \label{fig:branch}
\end{figure}

For $x\gg 0$ we will find contributions from all four asymptotic wave components and the expansion of the solution is
\begin{equation}    \label{H1right}
\phi^{(a)}(x) \stackrel{x\gg 0}{\sim} a_1^+ \phi_1^+(x) +a_2^+ \phi_2^+(x) +a_3^+ \phi_3^+(x) +a_4^+ \phi_4^+(x),
\end{equation}
for constant coefficients $a_n^+$. For large positive $x$ there are contributions from the two stationary-phase points (\ref{stat1}) and (\ref{stat3}). These points are the local wave-vectors of modes 2 and 3, respectively, in the far-right region, where they have the normalized asymptotic expressions (\ref{+2}) and (\ref{+3}). The steepest descent method gives the coefficients $a_2^+$ and $a_3^+$ of these modes:
\begin{align}
a_2^+=  &  (-1)^\frac{13}{8} \sqrt{\pi}  2^{\frac{1}{2}-\frac{ik_c}{2\alpha}}  e^{\frac{\pi k_c}{8\alpha}}  \alpha (\alpha k_c)^{\frac{i\omega}{\alpha}-\frac{ik_c}{4\alpha}-\frac{1}{4}},    \label{a2+}    \\
a_3^+=  & \frac{ (-1)^\frac{9}{8} 2 \pi  e^{\frac{\pi\omega}{\alpha} - \frac{\pi k_c}{8\alpha}}  \alpha (\alpha k_c)^{\frac{i\omega}{\alpha}+\frac{ik_c}{4\alpha}-\frac{1}{4}} }{\Gamma \left(  \frac{1}{2}+\frac{ik_c}{2\alpha} \right) } .  \label{a3+}  
\end{align}
There is also a contribution from the part of the contour $C'$ that runs around the singularity at $k=0$. For positive $x$ the integrand exponentially decreases along the positive imaginary $k$-axis (see top plot in Fig.~\ref{fig:Hermpos}) and we move the contour upwards so that it runs along both sides of the branch cut, as shown in Fig.~\ref{fig:branch}. The exponential decrease along the positive imaginary $k$-axis is very rapid for large $x$ so only the part of the contour in Fig.~\ref{fig:branch} close to $k=0$ is significant for the asymptotic integral. We can therefore compute this part of the integral by expanding the integrand around $k=0$. The Taylor expansion of the Hermite-function factor around $k=0$ is
\begin{align}
H_{-\frac{1}{2}-\frac{ik_c}{2\alpha}} & \left(\sqrt{\frac{i}{\alpha k_c}} \, k \right)  =   \frac{\sqrt{\pi} 2^{-\frac{1}{2}-\frac{ik_c}{2\alpha}} }{\Gamma \left(  \frac{3}{4}+\frac{ik_c}{4\alpha} \right) }  \nonumber \\
 & \qquad\qquad\quad  -  \frac{\sqrt{i\pi} 2^{\frac{1}{2}-\frac{ik_c}{2\alpha}} k}{\sqrt{\alpha k_c}\,\Gamma \left(  \frac{1}{4}+\frac{ik_c}{4\alpha} \right) } +O(k^2).   \label{Htaylor}
\end{align}
Using the first two terms in this series we compute the integral along the contour in Fig.~\ref{fig:branch} for  $x\gg 0$, where we can take the ends of the contour to run to positive imaginary infinity along both sides of the branch cut. We make the substitution $k=i s/x$ in the integral and expand the integrand to find the two leading-order terms for large $x$. This gives the integral
\begin{align}
\sqrt{\pi} 2^{-\frac{1}{2}-\frac{ik_c}{2\alpha}}   \int ds \,\left(\frac{is}{x}\right)^{\frac{i\omega}{\alpha}} 
&  e^{-s}    \left[  \frac{1}{s  \, \Gamma \left(  \frac{3}{4}+\frac{ik_c}{4\alpha} \right) }  \right.   \nonumber \\
 &  \left. + \frac{2\sqrt{-i}}{x \sqrt{\alpha k_c}\,  \Gamma \left(  \frac{1}{4}+\frac{ik_c}{4\alpha} \right) }   \right] ,   \label{sintH}
\end{align}
where our original choice of contour and branch cut in the complex $k$-plane means that in (\ref{sintH}) the branch cut in the integrand runs along the positive real $s$-axis and the contour runs in from infinity above the branch cut, around $s=0$, then out to infinity below the branch cut. Both terms in the integral (\ref{sintH}) are related to Hankel's integral representation of the gamma function~\cite{whittaker}, which states
\begin{equation}  \label{Han}
\Gamma(z)=\frac{i}{2\sin(\pi z)} \int_{C_\Gamma} dt \, (-t)^{z-1} e^{-t}, \quad z\notin \text{Integers},
\end{equation}
where the branch cut in the integrand is along the positive real $t$-axis and the contour $C_\Gamma$ runs in from infinity above the branch cut, around $t=0$, then out to infinity below the branch cut. The branch cut of the log function is chosen differently in (\ref{sintH}) compared to (\ref{Han}), and we must bear this in mind when employing the latter. The result for (\ref{sintH}) follows from (\ref{Han}):
\begin{align}
-\sqrt{\pi} 2^{\frac{1}{2}-\frac{ik_c}{2\alpha}}  e^{\frac{\pi \omega}{2 \alpha}}  \Gamma \left( \frac{i \omega}{\alpha} \right)   &   \sinh\left( \frac{\pi \omega}{\alpha} \right)  x^{-\frac{i\omega}{\alpha}} 
    \left[  \frac{1}{ \Gamma \left(  \frac{3}{4}+\frac{ik_c}{4\alpha} \right) }  \right.   \nonumber \\
 &  \left. + \frac{2\sqrt{i} \,\omega}{x \alpha \sqrt{\alpha k_c}\,  \Gamma \left(  \frac{1}{4}+\frac{ik_c}{4\alpha} \right) }   \right] .   \label{Hlowk1}
\end{align}
Referring to (\ref{+1}) and (\ref{+4}), we see that (\ref{Hlowk1}) is a superposition of the two low-$k$ asymptotic wave components on the far right (modes 1 and 4). Solving for the coefficients in this superposition we find $a_1^+$ and $a_4^+$ in (\ref{H1right}):
\begin{align}
a_1^+ = - &  \frac{   \sqrt{\omega} \, e^{\frac{\pi \omega}{2 \alpha}}  \Gamma \left( \frac{i \omega}{\alpha} \right)   \sinh\left( \frac{\pi \omega}{\alpha} \right) }{  \sqrt{2} \,  \Gamma \left( \frac{1}{2}+ \frac{i k_c}{2 \alpha} \right)  }
    \left[    \Gamma \left(  \frac{1}{4}+\frac{ik_c}{4\alpha} \right)  \right.   \nonumber \\
 &   \qquad\qquad\qquad  \left. + 2i \sqrt{\frac{i \alpha}{ k_c}}\ \Gamma \left(  \frac{3}{4}+\frac{ik_c}{4\alpha} \right)   \right] ,   \label{a1+}  \\
 a_4^+ = - &  \frac{   \sqrt{\omega} \, e^{\frac{\pi \omega}{2 \alpha}}  \Gamma \left( \frac{i \omega}{\alpha} \right)   \sinh\left( \frac{\pi \omega}{\alpha} \right) }{  \sqrt{2} \,  \Gamma \left( \frac{1}{2}+ \frac{i k_c}{2 \alpha} \right)  }
    \left[    \Gamma \left(  \frac{1}{4}+\frac{ik_c}{4\alpha} \right)  \right.   \nonumber \\
 &   \qquad\qquad\qquad  \left. - 2i \sqrt{\frac{i \alpha}{ k_c}}\ \Gamma \left(  \frac{3}{4}+\frac{ik_c}{4\alpha} \right)   \right] .   \label{a4+}
\end{align}
Here the expressions have been simplified a little by use of Legendre's duplication formula~\cite{htf1} for the gamma function, which gives
\begin{equation}
 \Gamma \left(  \frac{1}{2}+\frac{ik_c}{2\alpha} \right) = \frac{ 2^{- \frac{1}{2}+\frac{ik_c}{2\alpha}}  }{ \sqrt{\pi}  }   \Gamma \left(  \frac{1}{4}+\frac{ik_c}{4\alpha} \right)   \Gamma \left(  \frac{3}{4}+\frac{ik_c}{4\alpha} \right) .
\end{equation}

This completes the decomposition of the solution $\phi^{(a)}(x)$ into the asymptotic modes on both sides of the flow. The content of this solution in terms of mode scattering can be seen from (\ref{H1left}) and (\ref{H1right}), together with the lower ray plot in Fig.~\ref{fig:rays}. The input wave is incident from the right and is a superposition of the low-$k$ modes 1 and 4 with different amplitudes (\ref{a1+}) and (\ref{a4+}) (see Fig.~\ref{fig:rays}). Mode 1 propagates through the horizon regions and continues to the left, while mode 4 reverses its group velocity and is transformed to mode 3, as in the ray plots in Fig.~\ref{fig:rays}. In addition, there is scattering of the input modes 1 and 4 into mode 2 on the right. In fact there is also scattering of input mode 1 into mode 3 on the right and of input mode 4 into mode 1 on the left, but this will become clearer when we calculate the scattering coefficients. Note that there is no scattering into mode 4 on the left. Although both of the incident modes 1 and 4 from the right scatter into mode 4 on the left, the weighting of the incident modes in this solution causes cancellation of the scattering into mode 4 on the left.

The norm flux of any monochromatic solution is constant throughout the flow (see Appendix~\ref{app:conserved}). It must therefore be the case that the norm flux of the asymptotic wave decomposition (\ref{H1left}) on the far left of the flow is equal to that of the decomposition (\ref{H1right}) on the far right. As discussed in Appendix~\ref{app:asymptotic}, the norm flux of the normalized asymptotic wave components is $\pm1$, with a sign given by the product of the signs of the co-moving frequency and the group velocity. Also, a superposition of asymptotic wave components has a norm flux which is the sum of the fluxes of the individual components. For the solution (\ref{H1left}) and (\ref{H1right}), constancy of the norm flux therefore implies
\begin{equation}
-\left| a_1^- \right|^2 = - \left| a_1^+ \right|^2 + \left| a_2^+ \right|^2 -  \left| a_3^+ \right|^2 + \left| a_4^+ \right|^2.
\end{equation}
One can verify that this relation indeed holds for the coefficients (\ref{a1-}), (\ref{a2+}), (\ref{a3+}), (\ref{a1+}), and (\ref{a4+}). The calculation of the absolute values of the coefficients requires the gamma-function identities~\cite{htf1}
\begin{align}
\Gamma(z)\Gamma(-z) = & \, - \frac{\pi}{z\sin(\pi z)},  \label{gammaid1} \\
\Gamma(z)\Gamma(1-z) = & \,  \frac{\pi}{\sin(\pi z)}.  \label{gammaid2}
\end{align}

\subsubsection{Second solution: branch cut along negative imaginary $k$-axis}
We now take the branch cut in the integrand to lie along the negative imaginary $k$-axis (Fig.~\ref{fig:Hermneg}), with the solution defined by the contour $C''$. We denote this solution by $\phi^{(b)}(x)$ and calculate its two expansions in terms of the asymptotic wave components, one for $x\ll 0$ and one for $x\gg 0$. 
 
 For $x\ll 0$ we have a stationary-phase contribution to the integral from the point (\ref{stat2}), as in the first solution. But now there is also a branch-cut contribution because in deforming the contour $C''$ into the region along the negative imaginary axis (where the integrand decreases exponentially) it gets wrapped around the singularity at $k=0$. This last contribution will give low-$k$ wave components corresponding to modes 2 and 3 on the far left. The decomposition of $\phi^{(b)}(x)$ for $x\ll 0$ will consequently have three asymptotic wave components:
 \begin{equation}    \label{H2left}
\phi^{(b)}(x) \stackrel{x\ll 0}{\sim} b_1^- \phi_1^-(x) +b_2^- \phi_2^-(x) +b_3^- \phi_3^-(x).
\end{equation}
The calculation of $b_1^-$ by the steepest descent method is almost identical to that of $a_1^-$ in the first solution. The different position of the branch cut in the two cases means that  $b_1^-$ differs from $a_1^-$ by a simple factor:
\begin{equation}  \label{b1-}
b_1^- =  e^{-\frac{2\pi\omega}{\alpha}}  a_1^-.
\end{equation}
The branch-cut contribution gives the coefficients $b_2^-$ and $b_3^-$, and the calculation here has only minor differences from that represented in Fig.~\ref{fig:branch}. We again use the Taylor expansion (\ref{Htaylor}) of the Hermite function and Hankel's integral (\ref{Han}). The final results are simply related to the low-$k$ coefficients (\ref{a1+}) and (\ref{a4+}) in the previous solution:
\begin{equation}  \label{b2-}
b_2^- = - e^{-\frac{\pi\omega}{\alpha}}  a_4^+,  \qquad b_3^- = - e^{-\frac{\pi\omega}{\alpha}}  a_1^+.
\end{equation}

For $x\gg 0$ there is no branch-cut contribution to the integral as the part of the contour $C''$ near $k=0$ can be moved upwards along the positive imaginary axis where the integrand is exponentially small (see Fig.~\ref{fig:Hermneg}). There are stationary-phase contributions from the points (\ref{stat1}) and (\ref{stat3}), as in the previous solution, but here the new position of the branch cut in the integrand gives some minor differences. We obtain an expansion
\begin{equation}    \label{H2right}
\phi^{(b)}(x) \stackrel{x\gg 0}{\sim} b_2^+ \phi_2^+(x) +b_3^+ \phi_3^+(x)
\end{equation}
in which the coefficients $b_2^+$ and $b_3^+$ are simply related to (\ref{a2+}) and (\ref{a3+}):
\begin{equation}  \label{b2+}
b_2^+ =  a_2^+,  \qquad b_3^+ =  e^{-\frac{2\pi\omega}{\alpha}}  a_3^+.
\end{equation}

The expansions (\ref{H2left}) and (\ref{H2right}) reveal how the modes scatter in the solution $\phi^{(b)}(x)$. Referring to Fig.~\ref{fig:rays}, we see that the incident wave is a superposition of modes 2 and 3 from the left. Mode 2 propagates through into the right-hand region and there is also scattering into mode 3 on the right and mode 1 on the left. Note that there is no mode-4 component on the left even though the incident mode-3 component on the left would normally convert to a blue-shifted mode-4 wave on the same side. As in the solution $\phi^{(a)}(x)$, here the different contributions to mode 4 on the left exactly cancel.

The constancy of the norm flux for $\phi^{(b)}(x)$ implies the following relation between the ``$b$" coefficients for the modes on the left and the right:
\begin{equation}
-\left| b_1^- \right|^2 + \left| b_2^- \right|^2 - \left| b_3^- \right|^2 =  \left| b_2^+ \right|^2 -  \left| b_3^+ \right|^2.
\end{equation}
The coefficients (\ref{b1-}), (\ref{b2-}), and (\ref{b2+}) indeed satisfy this relation.

\subsection{Two solutions with $_1\!F_1(a;b;z)$ in their integral representations}
\begin{figure}[!htbp]
\includegraphics[width=\linewidth]{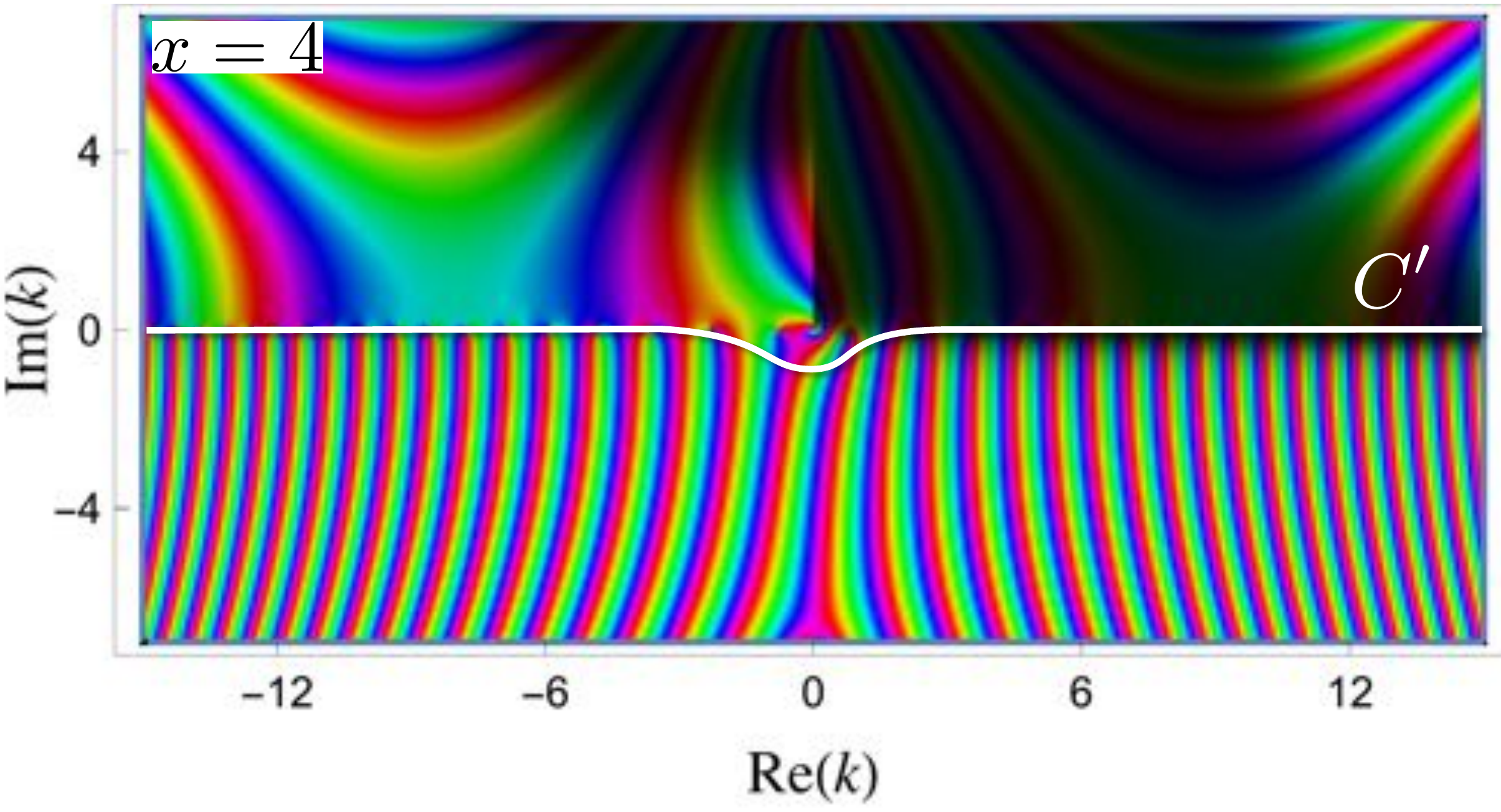}   
\includegraphics[width=\linewidth]{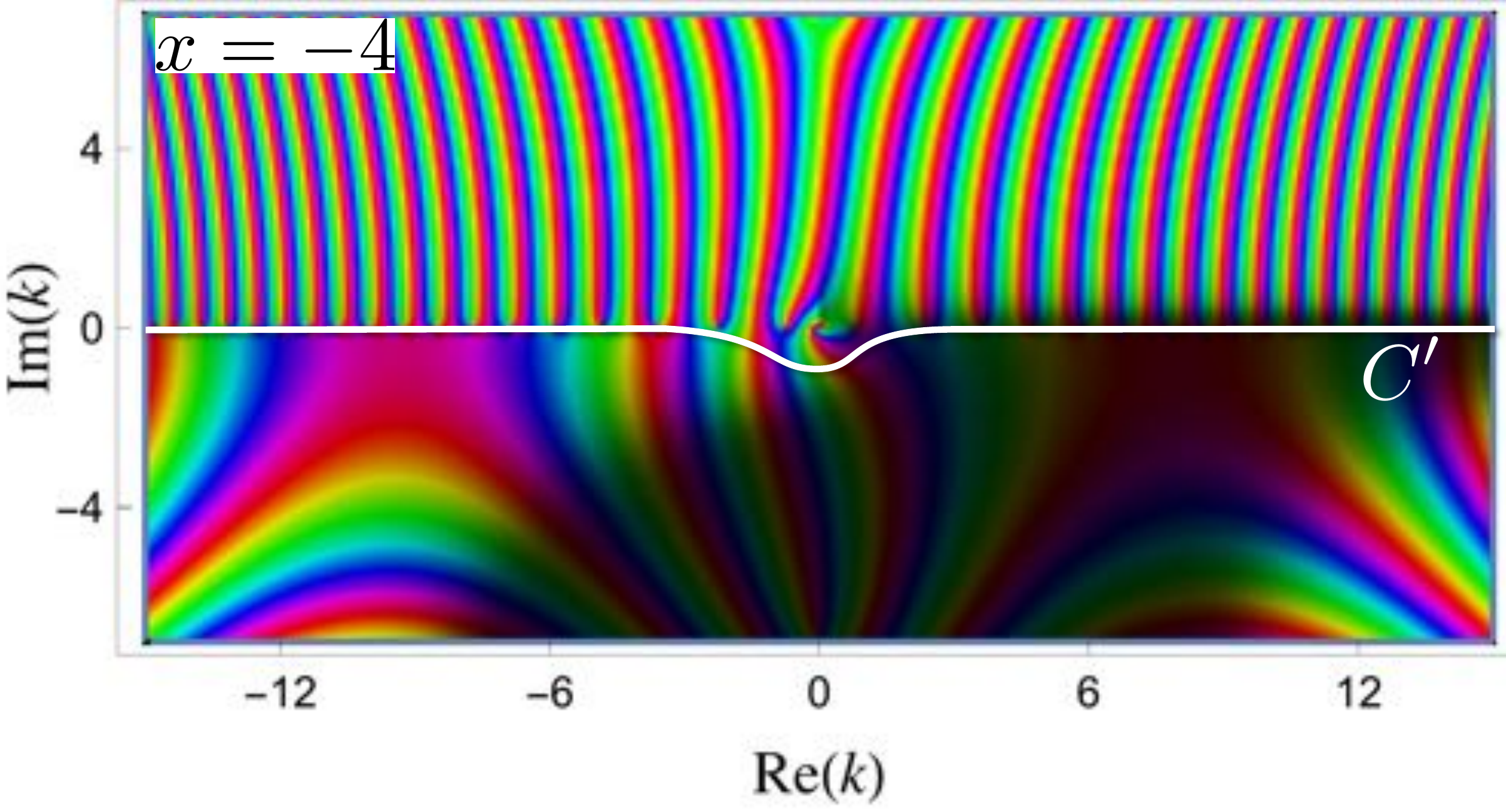}   
\caption{The integrand in (\ref{four}) plotted in the complex $k$-plane, with $c_1=0$ and $c_2=1$ in (\ref{ksol}). Colour indicates the phase while brightness indicates the absolute value (brighter is bigger). The branch cut is placed along the positive imaginary axis. Parameter values are $\omega=1$, $\alpha=1/2$ and $k_c=5$. The top plot is for $x=4$, the bottom for $x=-4$. With $C'$ as the contour in (\ref{four}) this defines an exact solution of (\ref{mono}).
} \label{fig:Hyperpos}
\end{figure}

\begin{figure}[!htbp]
\includegraphics[width=\linewidth]{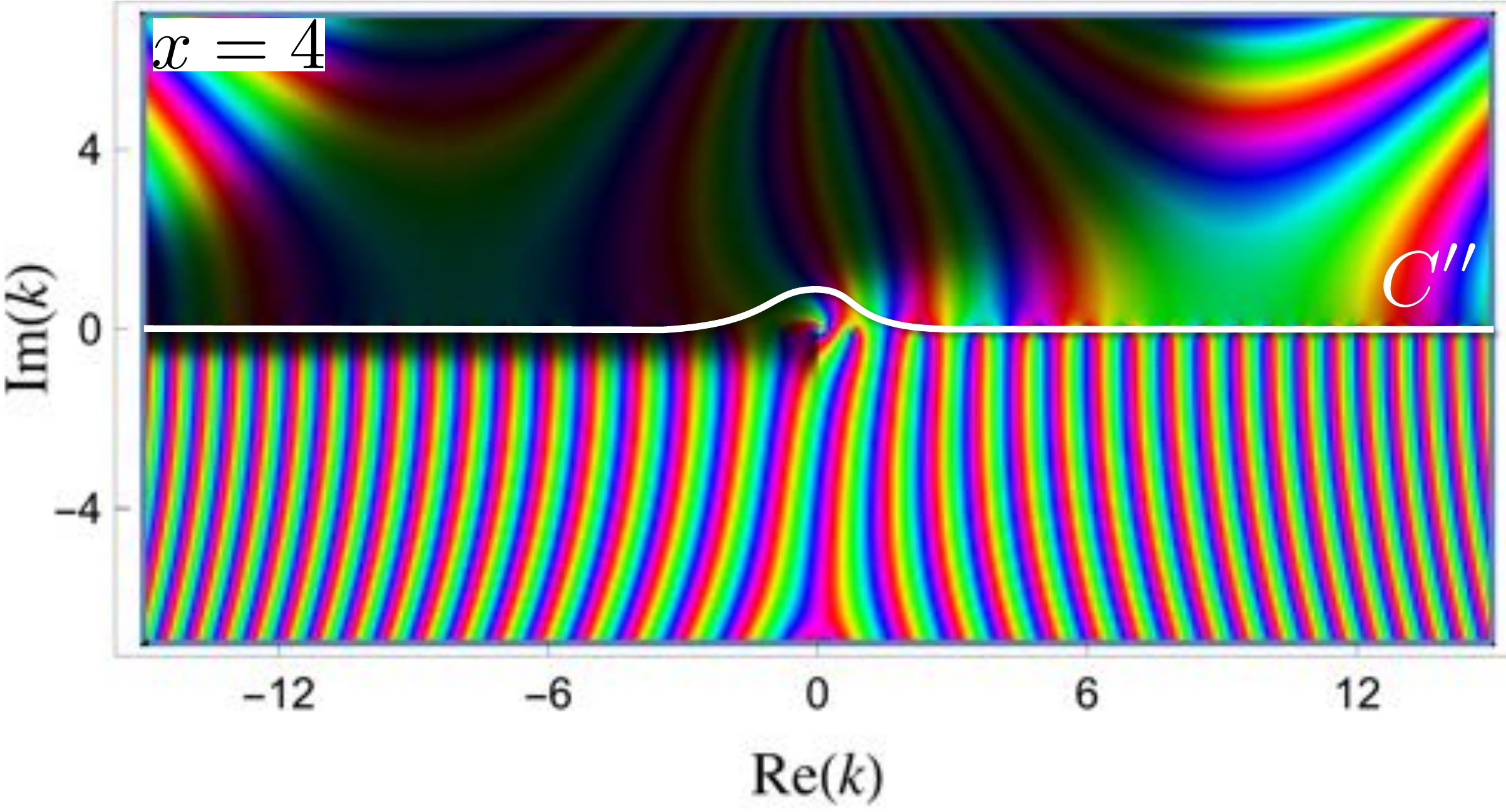}   
\includegraphics[width=\linewidth]{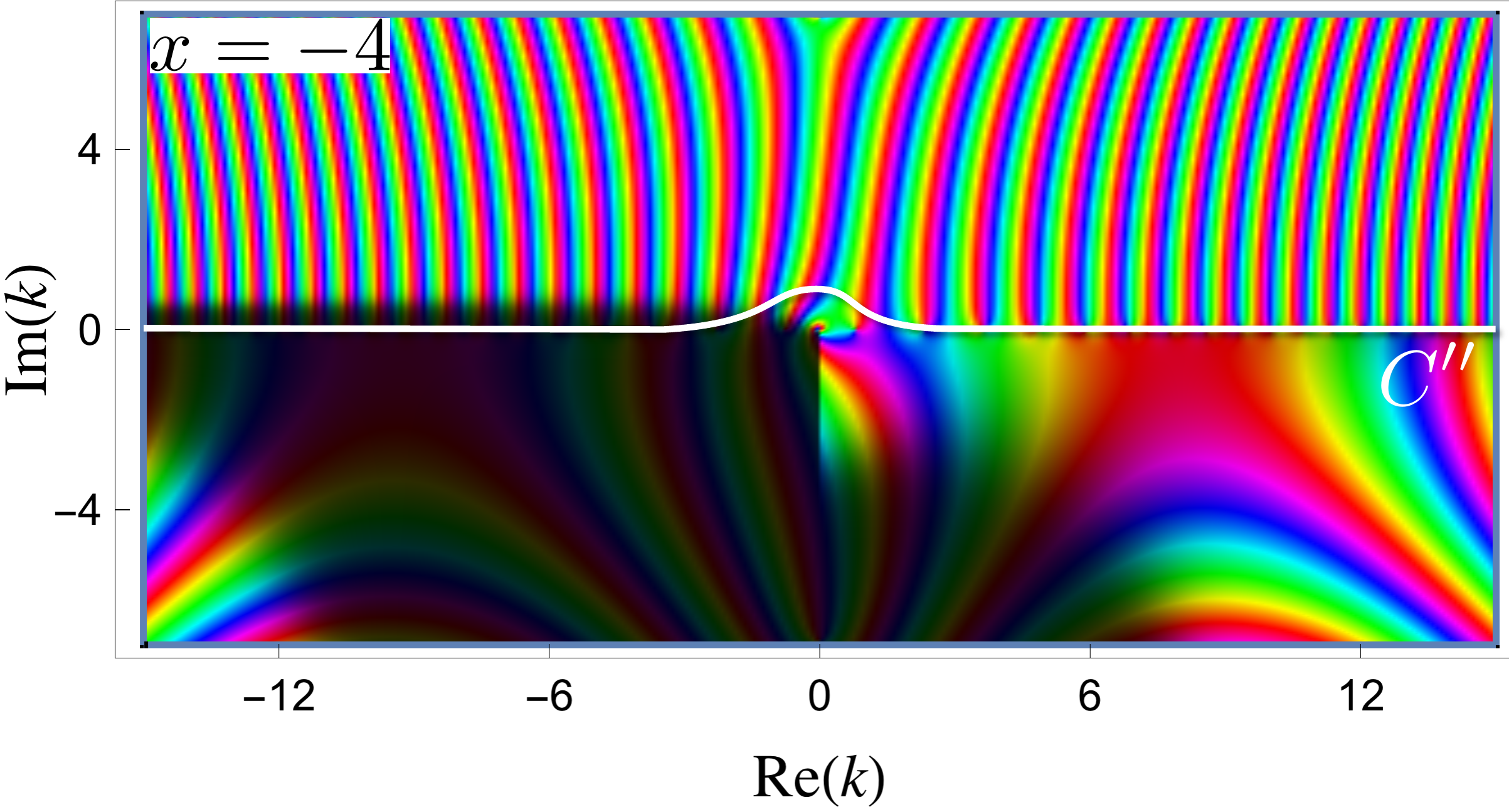}   
\caption{The same as Fig.~\ref{fig:Hyperpos}, but with the branch cut along the negative imaginary axis and a different contour $C''$. This defines another independent solution of (\ref{mono}).
} \label{fig:Hyperneg}
\end{figure}

We now take $c_1=0$ and $c_2=1$ in (\ref{ksol}), so that the integrand in (\ref{four}) contains the confluent hypergeometric  function $_1\!F_1(a;b;z)$. This integrand is plotted in the complex $k$-plane in Figs.~\ref{fig:Hyperpos} and~\ref{fig:Hyperneg}, with the branch cut running along the positive or negative imaginary axis, respectively. A solution of (\ref{mono}) is obtained for each choice of branch cut and corresponding contour $C'$ or $C''$. We proceed to find the expansions of these two solutions in terms of the asymptotic waves. The derivation follows exactly the steps described above when the integrand contained the Hermite function.

Figures~\ref{fig:Hyperpos} and~\ref{fig:Hyperneg} indicate that there are points of stationary phase on the real $k$-axis, but only for parts of the integrand. In the region near the positive real $k$-axis, the integrand appears to be the sum of a part with a stationary-phase point and a part with no extrema of the phase, and similarly for the region near the negative real $k$-axis. These stationary-phase features in the plots are present for both positive and negative $x$ and they move out along the real $k$-axis as $|x|$ increases. We  confirm this picture by looking at the asymptotic expansion of the hypergeometric function in (\ref{ksol}) for $k\gg0$ and $k\ll0$. We refer to~\cite{dlmf} for asymptotic expansions of the hypergeometric function for large variable. In our case the form $ik^2/(\alpha k_c)$ of the variable means there is an asymptotic expansion that is valid for both $k\gg0$ and $k\ll0$, which to leading order is
\begin{align}
 & _1\!F_1\left(\frac{1}{4}+\frac{ik_c}{4\alpha};\frac{1}{2};\frac{ik^2}{\alpha k_c}  \right)   \sim    \frac{\sqrt{i\pi} \,e^{-\frac{\pi k_c}{4\alpha}}}{\Gamma\left( \frac{1}{4}-\frac{ik_c}{4\alpha}  \right)}   \left( \frac{ik^2}{\alpha k_c} \right)^{ -\frac{1}{4}-\frac{ik_c}{4\alpha} } 
 \nonumber   \\
& \qquad  \quad\ + \frac{\sqrt{\pi}}{\Gamma\left( \frac{1}{4}+\frac{ik_c}{4\alpha}  \right)}   \left( \frac{ik^2}{\alpha k_c} \right)^{ -\frac{1}{4}+\frac{ik_c}{4\alpha} }  \exp\left( \frac{i k^2}{\alpha k_c} \right) ,  \nonumber   \\
& \qquad\quad   -\frac{\pi}{2} < \arg(k) <  \frac{\pi}{2}\ \ \text{or}\  \ \frac{\pi}{2} < \arg(k) <  \frac{3\pi}{2}.  \label{Fk}
\end{align}
The two terms in the expansion (\ref{Fk}) give terms with different phase factors in the integrand in  (\ref{four}) (we take just the hypergeometric-function term in (\ref{ksol})). The first term in the resulting integrand has a phase factor $\exp\left[ikx- ik^2/(2\alpha k_c) \right]$, which has a point of stationary phase at
\begin{equation} \label{stat4}
k=\alpha k_c x.
\end{equation}
As (\ref{Fk}) is valid for both $k\gg0$ and $k\ll0$, the first term in the integrand thus has a stationary-phase point (\ref{stat4}) on the positive real $k$-axis for $x\gg0$ and on the negative real $k$-axis for $x\ll0$. The second term in (\ref{Fk}) gives a term in the integrand in (\ref{four}) with a phase factor $\exp\left[ikx+ ik^2/(2\alpha k_c) \right]$ and a point of stationary phase at
\begin{equation} \label{stat5}
k= -\alpha k_c x.
\end{equation}
The second term in the integrand thus has a stationary-phase point (\ref{stat5}) on the negative real $k$-axis for $x\gg0$ and on the positive real $k$-axis for $x\ll0$. These results confirm what is already apparent from the plots in Figs.~\ref{fig:Hyperpos} and~\ref{fig:Hyperneg}.

The asymptotics (\ref{Fk}) is not valid on the imaginary $k$-axis, and we have not given an expansion of the hypergeometric function valid in this region of the complex $k$-plane. We do not include it here because it does not give any points of stationary phase for terms in the integrand in (\ref{four}).

The stationary-phase points (\ref{stat4}) and (\ref{stat5}) lie on both contours $C'$ and $C''$ and give contributions to the asymptotic expansions of the integral  (\ref{four}) for large $|x|$. These contributions are calculated by the method of steepest descent. There may also be a branch-cut contribution, which is evaluated as above for the solutions generated by the Hermite function.

\subsubsection{First solution: branch cut along positive imaginary $k$-axis}
With the branch cut in the integrand chosen to lie along the positive imaginary $k$-axis (Fig.~\ref{fig:Hyperpos}) the solution is defined by the contour $C'$. We denote this solution by $\phi^{(c)}(x)$ and find its expansions in terms of the asymptotic waves for $x\ll 0$ and for $x\gg 0$. 

For $x\ll0$ there are stationary-phase contributions from the points (\ref{stat4}) and (\ref{stat5}), which give asymptotic waves corresponding to modes 1 and 4, respectively. There is no branch-cut contribution and so the expansion is
\begin{equation}
\phi^{(c)}(x) \stackrel{x\ll 0}{\sim} c_1^- \phi_1^-(x) + c_4^- \phi_4^-(x) .    \label{F1left} 
\end{equation}
The steepest descent method gives the coefficients $c_1^-$ and $c_4^-$:

\begin{align}
c_1^-=  &  -\frac{2\pi  (-1)^\frac{7}{8}   e^{\frac{\pi\omega}{\alpha}-\frac{\pi k_c}{8\alpha}}   \alpha (\alpha k_c)^{\frac{i\omega}{\alpha}-\frac{ik_c}{4\alpha}-\frac{1}{4}}  }{  \Gamma\left( \frac{1}{4}-\frac{ik_c}{4\alpha}  \right)  }  ,  \label{c1-}   \\
c_4^-=  &  -\frac{2\pi  (-1)^\frac{1}{8}   e^{-\frac{\pi k_c}{8\alpha}}   \alpha (\alpha k_c)^{\frac{i\omega}{\alpha}+\frac{ik_c}{4\alpha}-\frac{1}{4}}  }{  \Gamma\left( \frac{1}{4}+\frac{ik_c}{4\alpha}  \right)  }  .  \label{c4-} 
\end{align}

For $x\gg 0$, the solution $\phi^{(c)}(x)$ has contributions from all four asymptotic wave components:
\begin{equation}
\phi^{(c)}(x) \stackrel{x\gg 0}{\sim} c_1^+ \phi_1^+(x) +   c_2^+ \phi_2^+(x)  +   c_3^+ \phi_3^+(x) + c_4^+ \phi_4^+(x) .    \label{F1right} 
\end{equation}
The two stationary-phase points (\ref{stat4}) and (\ref{stat5}) give rise to the mode-2 and mode-3 waves on the far right, with coefficients
\begin{align}
c_2^+=  &  -\frac{2\pi  (-1)^\frac{7}{8}   e^{-\frac{\pi k_c}{8\alpha}}   \alpha (\alpha k_c)^{\frac{i\omega}{\alpha}-\frac{ik_c}{4\alpha}-\frac{1}{4}}  }{  \Gamma\left( \frac{1}{4}-\frac{ik_c}{4\alpha}  \right)  }  ,  \label{c2+}   \\
c_3^+ =  &  -\frac{2\pi  (-1)^\frac{1}{8}   e^{\frac{\pi\omega}{\alpha} -\frac{\pi k_c}{8\alpha}}   \alpha (\alpha k_c)^{\frac{i\omega}{\alpha}+\frac{ik_c}{4\alpha}-\frac{1}{4}}  }{  \Gamma\left( \frac{1}{4}+\frac{ik_c}{4\alpha}  \right)  }  .  \label{c3+} 
\end{align}
There is also a branch-cut contribution, which gives the low-$k$ asymptotic wave components on the right (modes 1 and 4). The relevant contour is as in Fig.~\ref{fig:branch}, but here the integrand is different. As before, we only require the integrand in the region near $k=0$ in order to evaluate the branch-cut contribution. We employ the Taylor expansion of the hypergeometric function around $k=0$:
\begin{equation}
 _1\!F_1\left(\frac{1}{4}+\frac{ik_c}{4\alpha};\frac{1}{2};\frac{ik^2}{\alpha k_c}  \right)  =  1- \frac{ (k_c-i\alpha)k^2 }{ 2\alpha^2k_c  }  +O(k^4). \label{Ftaylor}
\end{equation}
As previously, we make the substitution $k=i s/x$ in the resulting integral and expand the integrand to find the first two leading-order terms for large $x$. This gives the integral
\begin{equation}
  \int ds \,\left(\frac{is}{x}\right)^{\frac{i\omega}{\alpha}} 
  e^{-s}    \left[  \frac{1}{s  }  
    + \frac{ s }{ 2 x^2 \alpha^2 }   \right] ,   \label{sintF}
\end{equation}
where the branch cut in the integrand runs along the positive real $s$-axis and the contour runs in from infinity above the branch cut, around $s=0$, then out to infinity below the branch cut. Evaluation of (\ref{sintF}) using Hankel's integral (\ref{Han}) produces
\begin{align}
- 2  e^{\frac{\pi \omega}{2 \alpha}}  \Gamma \left( \frac{i \omega}{\alpha} \right)   &   \sinh\left( \frac{\pi \omega}{\alpha} \right)  x^{-\frac{i\omega}{\alpha}} 
    \left[ 1   + \frac{ i \omega ( \alpha  +i \omega)  }{2 x^2 \alpha^4  }   \right] .   \label{Flowk1}
\end{align}
This is a superposition of the two low-$k$ asymptotic wave components (\ref{+1}) and (\ref{+4}), but note that (\ref{Flowk1}) does not have a term falling off as $1/x$, whereas both  (\ref{+1}) and (\ref{+4}) contain such a term. It must therefore be the case that (\ref{Flowk1}) is an equal superposition of  (\ref{+1}) and (\ref{+4}), so that the terms containing $x^{-i\omega/\alpha-1}$ cancel out. This would require the ratio of the two terms in (\ref{Flowk1}) to be equal to the ratio of the first and third terms in (\ref{+1}) and (\ref{+4}), as is indeed the case. Solving for the coefficients $c_1^+$ and $c_4^+$ in the superposition we obtain
\begin{equation}
c_1^+=c_4^+  = - \sqrt{2\omega} \, e^{\frac{\pi \omega}{2 \alpha}}  \Gamma \left( \frac{i \omega}{\alpha} \right)     \sinh\left( \frac{\pi \omega}{\alpha} \right)  .   \label{c1+} 
\end{equation}

The expansions (\ref{F1left}) and (\ref{F1right}), together with (\ref{c1+}), show that the solution $\phi^{(c)}(x)$ is an incident wave from the right that is an equal superposition of modes 1 and 4 (see Fig.~\ref{fig:rays}). These incident modes scatter into all outgoing modes on the left and right. The ``$c$" coefficients obey the relation that follows from constancy of the norm flux:
\begin{equation}
-\left| c_1^- \right|^2+\left| c_4^- \right|^2 = - \left| c_1^+ \right|^2 + \left| c_2^+ \right|^2 -  \left| c_3^+ \right|^2 + \left| c_4^+ \right|^2.
\end{equation}

\subsubsection{Second solution: branch cut along negative imaginary $k$-axis}
For the second  solution generated by the hypergeometric function, we take the branch cut in the integrand to lie along the negative imaginary $k$-axis (Fig.~\ref{fig:Hyperneg}). The solution is defined by the contour $C''$ and we denote it by $\phi^{(d)}(x)$. The calculation of the expansions of the solution in terms of the asymptotic wave components has by now been well rehearsed and here we quote the results.

For $x\ll0$ the expansion contains all four asymptotic wave components:
 \begin{equation}    \label{F2left}
\phi^{(d)}(x) \stackrel{x\ll 0}{\sim} d_1^- \phi_1^-(x) +d_2^- \phi_2^-(x) +d_3^- \phi_3^-(x) +d_4^- \phi_4^-(x),
\end{equation}
where the coefficients are given by
\begin{equation}
d_1^- = e^{-\frac{2 \pi\omega}{\alpha}} c_1^-,  \quad  d_2^- = d_3^- = - e^{-\frac{\pi\omega}{\alpha}}  c_1^+,   \quad 
d_4^- = c_4^-.
\end{equation}
For $x\gg0$ the expansion has two wave components, for modes 2 and 3:
 \begin{equation}    \label{F2right}
\phi^{(d)}(x) \stackrel{x\gg 0}{\sim}   d_2^+ \phi_2^+(x) +d_3^+ \phi_3^+(x) ,
\end{equation}
with coefficients
\begin{gather}
d_2^+ =  c_2^+,  \qquad  d_3^+ =   e^{-\frac{2\pi\omega}{\alpha}}  c_3^+ .
\end{gather}
The ``$d$" coefficients obey the norm-flux constancy condition
\begin{equation}
-\left| d_1^- \right|^2 + \left| d_2^- \right|^2 - \left| d_3^- \right|^2 + \left| d_4^- \right|^2 =  \left| d_2^+ \right|^2 -  \left| d_3^+ \right|^2.
\end{equation}

The solution $\phi^{(d)}(x)$ is an incident wave from the left that is an equal superposition of modes 2 and 3 (see Fig.~\ref{fig:rays}). The incident wave is scattered into all outgoing modes on the left and right. 

\section{Scattering coefficients}    \label{sec:scat}
The four independent solutions derived in the previous section contain all the information about scattering in the flow. To compute the scattering coefficients however, it is convenient to have solutions that contain just one incident mode. These are straightforwardly obtained by superposing our four solutions in an appropriate manner.

\subsection{Scattering of mode 2 incident from the left}   
\begin{figure}[!htbp]
\includegraphics[width=\linewidth]{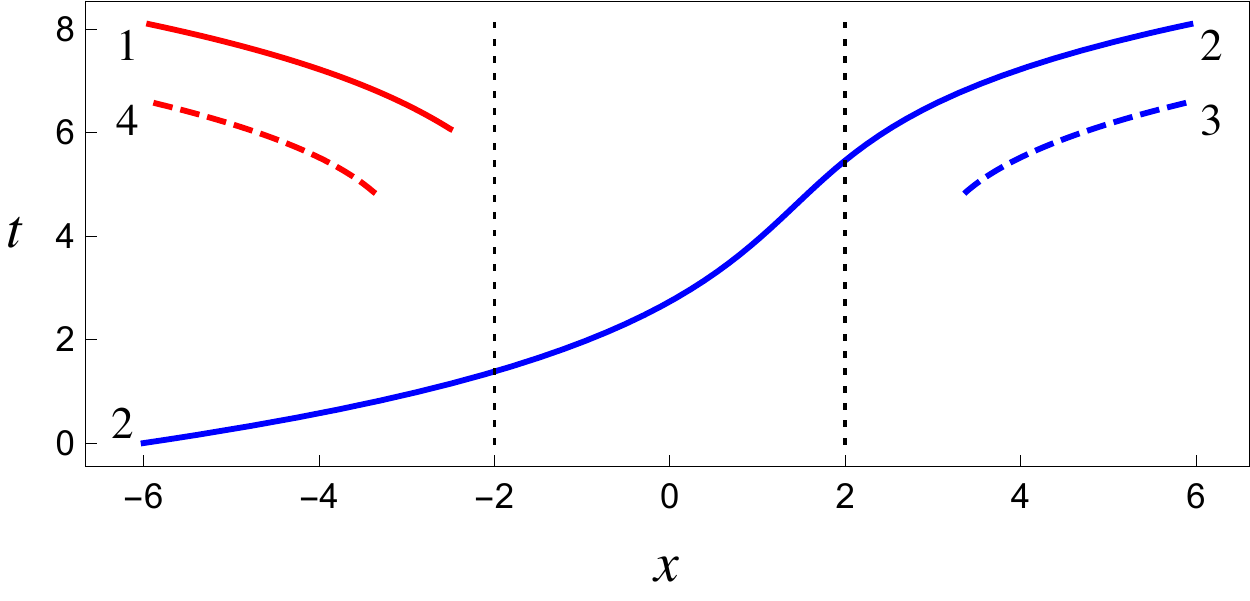}  
\caption{ Heuristic ray picture for the wave solution (\ref{2in}), whose only incident asymptotic wave component is the normalized mode-2 wave (\ref{-2}) on the left. The incident wave scatters into all outgoing modes, some of which have negative norm (modes 3 and 4) and some propagate to the left relative to the fluid (modes 1 and 4). As in Fig,~\ref{fig:rays}, the flow corresponds to a white-hole binary. Reversing $t$ (i.e. reading the plot top to bottom) corresponds to a black-hole binary, in which several modes in the past combine to give an outgoing low-$k$ mode 2 on the left.
} \label{fig:2in}
\end{figure}

We first construct a solution that corresponds to mode 2 incident from the left (see Fig.~\ref{fig:2in}). In this solution the asymptotic wave components on the far left do not include mode 3, and on the far right there are no asymptotic waves for modes 1 and 4. We also choose the mode-2 component on the left to be normalized so that it is exactly (\ref{-2}). Two of our four solutions have a wave incident from the left only, namely $\phi^{(b)}(x)$ and $\phi^{(d)}(x)$. We superpose these solutions so that on the far left the incident mode-3 component is removed and the mode-2 component is normalized. From (\ref{H2left}) and (\ref{F2left}) the required wave is
\begin{equation}   \label{2in}
\phi^{\text{2in}}(x)=  \frac{  d_3^- \phi^{(b)}(x)   -  b_3^-   \phi^{(d)}(x)  }{  d_3^- b_2^- -   b_3^- d_2^-}.
\end{equation}

A qualitative picture of the solution (\ref{2in}) is given in  Fig.~\ref{fig:2in}. The incident mode 2  is right-moving relative to the fluid and has positive norm. This incident mode propagates through to the right where it scatters into mode 3, which is also a right-mover but has negative norm. As a result of the scattering, the outgoing mode 2 on the right has been amplified (otherwise norm would not be conserved). This part of the scattering process occurs even in the limit of no dispersion, where it corresponds exactly to the Hawking effect at an event horizon. Because of dispersion there is also \emph{reflection} of mode 2 into modes which are left-moving relative to the fluid (modes 1 and 4 on the left). Moreover the reflection also involves further scattering into a negative-norm mode, namely mode 4 on the left, and therefore additional amplification of positive-norm components.

The scattering coefficients for this process are just the coefficients of the asymptotic wave components of (\ref{2in}) on the far left and far right. These are easily obtained from (\ref{H2left}), (\ref{H2right}), (\ref{F2left}), (\ref{F2right}) and (\ref{2in}). We denote the coefficient for scattering of incident mode 2 into outgoing mode $n$ by $S_{n,2}$, so that the expansions of (\ref{2in}) into asymptotic waves read
 \begin{align}   
\phi^{\text{2in}}(x) \stackrel{x\ll 0}{\sim}  & \, \phi_2^-(x) + S_{1,2} \phi_1^-(x) + S_{4,2} \phi_4^-(x), \label{2inleft} \\
\phi^{\text{2in}}(x) \stackrel{x\gg 0}{\sim}  & \,   S_{2,2} \phi_2^+(x)  +  S_{3,2} \phi_3^+(x). \label{2inright}
\end{align}
The most important information is given by the absolute values of the scattering coefficients. In computing $|S_{n,2}|^2$ we make use of the identities (\ref{gammaid1}) and (\ref{gammaid2}) to obtain
\begin{widetext}
 \begin{align}   
 \left| S_{1,2} \right|^2    =  & \, \left(  e^{ \frac{2\pi\omega}{\alpha} } -1  \right)^{-1}   \left[ - \frac{1}{2}     
     +    e^{ -\frac{\pi k_c}{4\alpha} }  \left(   \frac{ \pi \sqrt{ k_c }  } { 4 \sqrt{\alpha} }    \left| \Gamma\left( \frac{3}{4}+\frac{ik_c}{4\alpha}  \right)  \right|^{-2}   
   +  \frac{ \pi \sqrt{ \alpha }  } {  \sqrt{k_c} }  \left| \Gamma\left( \frac{1}{4}+\frac{ik_c}{4\alpha}  \right)  \right|^{-2}   \right)  \right]  ,    \label{S12}  \\
\left| S_{2,2} \right|^2    = & \,  \left(  1 - e^{ -\frac{2\pi\omega}{\alpha} } \right)^{-1}   \left[ \frac{1}{2}     
     +    e^{ -\frac{\pi k_c}{4\alpha} }  \left(   \frac{ \pi \sqrt{ k_c }  } { 4 \sqrt{\alpha} }    \left| \Gamma\left( \frac{3}{4}+\frac{ik_c}{4\alpha}  \right)  \right|^{-2}   
   +  \frac{ \pi \sqrt{ \alpha }  } {  \sqrt{k_c} }  \left| \Gamma\left( \frac{1}{4}+\frac{ik_c}{4\alpha}  \right)  \right|^{-2}   \right)  \right]   ,    \label{S22} \\
    \left| S_{3,2} \right|^2    =   & \, \left(  e^{ \frac{2\pi\omega}{\alpha} } -1  \right)^{-1}   \left[  \frac{1}{2}     
     +   e^{ -\frac{\pi k_c}{4\alpha} }  \left(   \frac{ \pi \sqrt{ k_c }  } { 4 \sqrt{\alpha} }    \left| \Gamma\left( \frac{3}{4}+\frac{ik_c}{4\alpha}  \right)  \right|^{-2}   
   +  \frac{ \pi \sqrt{ \alpha }  } {  \sqrt{k_c} }  \left| \Gamma\left( \frac{1}{4}+\frac{ik_c}{4\alpha}  \right)  \right|^{-2}   \right)  \right] ,    \label{S32}    \\
  \left| S_{4,2} \right|^2    =   & \, \left( 1-  e^{- \frac{2\pi\omega}{\alpha} }   \right)^{-1}   \left[ - \frac{1}{2}     
     +  e^{ -\frac{\pi k_c}{4\alpha} }  \left(   \frac{ \pi \sqrt{ k_c }  } { 4 \sqrt{\alpha} }    \left| \Gamma\left( \frac{3}{4}+\frac{ik_c}{4\alpha}  \right)  \right|^{-2}   
   +  \frac{ \pi \sqrt{ \alpha }  } {  \sqrt{k_c} }  \left| \Gamma\left( \frac{1}{4}+\frac{ik_c}{4\alpha}  \right)  \right|^{-2}   \right) \right] .     \label{S42} 
\end{align}
\end{widetext}
These are exact amplitudes of scattering coefficients for the linear flow profile. They were previously obtained by Busch and Parentani~\cite{bus12} by a different approach as part of a study of dispersive fields in de Sitter space. We refer to~\cite{bus12} for the application of these results to cosmological particle creation and black-hole thermodynamics when Lorentz invariance is broken.

 Similarly to the four wave solutions in the previous section, (\ref{2inleft}) and (\ref{2inright}) imply that $|S_{n,2}|^2$ are (up to a sign) equal to the norm fluxes of the outgoing modes. Constancy of the norm flux explains  the following identity:
\begin{equation}
1-  \left| S_{1,2} \right|^2  +   \left| S_{4,2} \right|^2 =  \left| S_{2,2} \right|^2  -  \left| S_{3,2} \right|^2 .
\end{equation}

The scattering amplitudes (\ref{S12})--(\ref{S42}) demonstrate the importance of dispersion in laboratory analogues of event horizons. Note the following simple relation that is clear from 
(\ref{S12})--(\ref{S42}):
\begin{equation}
\frac{  \left| S_{3,2} \right|  }{  \left| S_{2,2} \right| } = \frac{  \left| S_{1,2} \right|  }{  \left| S_{4,2} \right| }  = e^{ -\frac{\pi\omega}{\alpha} } .   \label{ratio}
\end{equation}
The first ratio in (\ref{ratio}) has the same value as in the non-dispersive case (see below) but dispersion leads to non-zero values for $S_{1,2}$ and $S_{4,2}$ whose ratio matches that of $S_{3,2}$ and $S_{2,2}$. An interesting question is whether the elementary relation (\ref{ratio}) is a property of the linear flow profile with \emph{arbitrary} anomalous dispersion, but our results allow no conclusions on this point.

The non-dispersive result for the scattering amplitudes is very simple and was first derived by Hawking~\cite{haw74}. To extract the non-dispersive case from (\ref{S12})--(\ref{S42})  we can employ asymptotic expansions of the gamma functions for large $k_c$. The following complete asymptotic expansion for large $z$ will allow us to take the required limit~\cite{htf1} :
\begin{align}
\ln \left[\Gamma(a+z) \right] \sim   &  \, \left(a+z-\frac{1}{2}\right) \ln z - z + \frac{1}{2} \ln (2\pi)  \nonumber \\
&+ \sum_{n=1}^{\infty}  \frac{ (-1)^{n+1} B_{n+1} (a) }{ n(n+1) z^n  } ,  \label{lnGamma}
\end{align}
where $B_n(z)$ are the Bernoulli polynomials. Employing this expansion in (\ref{S12})--(\ref{S42}) we can compute the leading-order scattering amplitudes as $k_c\to\infty$:
\begin{align}
 \left| S_{1,2} \right|^2    \sim  & \, \left(  e^{ \frac{2\pi\omega}{\alpha} } -1  \right)^{-1}  \frac{ \alpha^4 }{ 64 k_c^4 }  \left(  1+   \frac{5 \alpha^2 }{k_c^2 } \right)  ,  \label{S12as}  \\
\left| S_{2,2} \right|^2    \sim & \,  \left(  1 - e^{ -\frac{2\pi\omega}{\alpha} } \right)^{-1} \left[ 1+  \frac{ \alpha^4 }{ 64 k_c^4 }  \left(  1+   \frac{ 5\alpha^2 }{k_c^2 }  \right)  \right]  ,    \label{S22as} \\
    \left| S_{3,2} \right|^2    \sim   & \, \left(  e^{ \frac{2\pi\omega}{\alpha} } -1  \right)^{-1}  \left[ 1+  \frac{ \alpha^4 }{ 64 k_c^4 }  \left(  1+   \frac{5 \alpha^2 }{k_c^2 }  \right)  \right]  ,    \label{S32as}    \\ 
  \left| S_{4,2} \right|^2    \sim   & \,  \left(  1 - e^{ -\frac{2\pi\omega}{\alpha} } \right)^{-1}   \frac{ \alpha^4 }{ 64 k_c^4 }  \left(  1+   \frac{5 \alpha^2 }{k_c^2 } \right) .     \label{S42as} 
\end{align}
These expansions should be treated with caution. The appearance of exponentials containing $\pm k_c/\alpha$ in the asymptotic expansions of the gamma functions in (\ref{S12})--(\ref{S42}) means it is not possible to develop complete asymptotic expansions of the scattering amplitudes in powers of $\alpha/k_c$. For $k_c\to\infty$, we obtain from (\ref{S12as})--(\ref{S42as}) the Hawking result
\begin{gather}
 \left| S_{1,2} \right|^2    =   \left| S_{4,2} \right|^2  = 0 ,  \label{S12haw}  \\
\left| S_{2,2} \right|^2    =  \left(  1 - e^{ -\frac{2\pi\omega}{\alpha} } \right)^{-1} ,    \quad
    \left| S_{3,2} \right|^2    =   \left(  e^{ \frac{2\pi\omega}{\alpha} } -1  \right)^{-1}  ,    \label{S32haw}   
\end{gather}
in which there is no scattering of the incident mode into modes left-moving relative to the fluid. The squared amplitude $ \left| S_{3,2} \right|^2$ in (\ref{S32haw}) for scattering into the right-moving negative-norm mode has the form of the Planck distribution.

The frequency dependence of the scattering amplitudes (\ref{S12})--(\ref{S42}) factors out neatly from the dependence on dispersion, the latter being a function of $k_c/\alpha$. This factorization may be unique to the linear flow profile. In heuristic terms, each frequency in the linear profile experiences the same profile shape in the region around a blocking point (where the group velocity changes sign), even though the blocking point is at a different position in the flow for each frequency. For a curving flow profile each frequency will experience a different flow shape near a blocking point and the dependence of the scattering coefficients on frequency, profile shape and dispersion will be very complicated~\cite{rob12,mac09,leo12,fin12,cou12,mic14,rob14,euv15}. 

The fact that classical waves in the flow experience scattering into modes of opposite norm implies spontaneous emission in the quantum theory, provided an appropriate quantum description exists. The derivation of quantum emission from classical scattering into negative-norm modes is well described elsewhere~\cite{rob12,bro95,unr95} and is not repeated here. Recall that we chose the sign of $t$ so that the flow corresponds to a white-hole binary, whereas a black-hole binary is obtained by $t\to-t$. For the black-hole binary, the scattering of mode 2 given by the solution (\ref{2in}) occurs backward in time, so that time runs from the top to the bottom in Fig.~\ref{fig:2in}. In this case the positive-norm mode 2 on the left contains negative-norm components in the past. As a consequence, the annihilation operator for mode-2 quanta is a sum of annihilation and creation operators for modes in the past. If all modes were in their ground state in the past then mode 2 will now contain quanta and there will be emission of low-$k$ waves to the left (see Fig.~\ref{fig:2in}, reading top to bottom). The expectation value for the number of quanta in mode 2 is $ \left| S_{3,2} \right|^2 + \left| S_{4,2} \right|^2 $, and is thus determined by the scattering into negative-norm modes. In the dispersionless case this gives the familiar thermal spectrum of quanta, as is seen from (\ref{S12haw}) and (\ref{S32haw}). When dispersion is included the spectrum is obtained from  (\ref{S32}) and (\ref{S42}) and is no longer thermal. 

Note that $\left| S_{4,2} \right|^2$ does not go to zero at large frequencies, but rather approaches a value given by the quantity in square brackets in (\ref{S42}). This means the result for quantum emission does not vanish at large frequency, in contrast to the dispersionless case. But we cannot of course employ the model assumed here at arbitrarily high frequencies. The missing ingredient is dissipation. When waves propagate in any medium there is a limit to the size of the wavelength that can be supported and one manifestation of this is the loss of energy from the wave into the medium. This is very clear in the case of water waves where it is readily observed how dissipation increases as the wavelength decreases into the capillary-wave regime. The influence of dissipation on the Hawking effect has been explored in~\cite{ada13,rob15}. 

\subsection{Scattering of mode 3 incident from the left}   
\begin{figure}[!htbp]
\includegraphics[width=\linewidth]{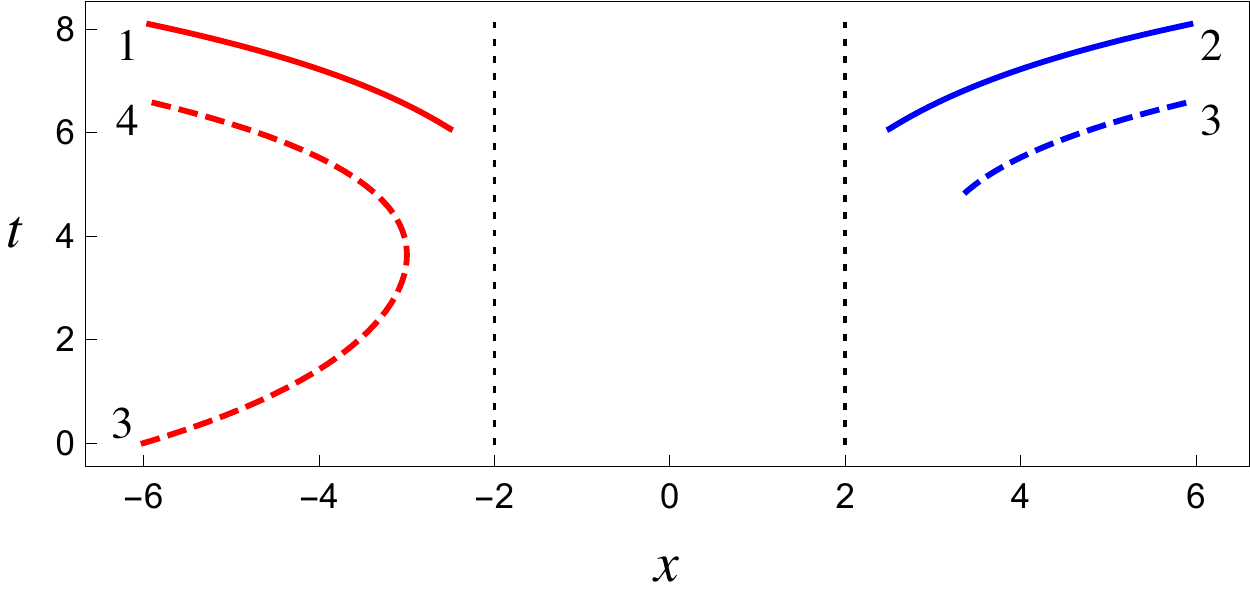}  
\caption{ Heuristic ray picture for the wave solution (\ref{3in}), whose only incident asymptotic wave component is the normalized mode-3 wave (\ref{-3}) on the left. The incident wave has negative norm and scatters into all outgoing modes, some of which have positive norm (modes 1 and 2) and some propagate to the right relative to the fluid (modes 2 and 3 on the right).
} \label{fig:3in}
\end{figure}

We now construct a solution that corresponds to mode 3 incident from the left (see Fig.~\ref{fig:3in}). Here the asymptotic wave components on the far left do not include mode 2, and on the far right there are no asymptotic waves for modes 1 and 4. The mode-3 component on the left is normalized to be exactly (\ref{-3}). This solution is given by a superposition of $\phi^{(b)}(x)$ and $\phi^{(d)}(x)$ that removes the mode-2 wave on the left and normalizes the mode-3 component on the left. From (\ref{H2left}) and (\ref{F2left}) we find the required wave:
\begin{equation}   \label{3in}
\phi^{\text{3in}}(x)=  \frac{  d_2^- \phi^{(b)}(x)   -  b_2^-   \phi^{(d)}(x)  }{  d_2^- b_3^- -   b_2^- d_3^-}.
\end{equation}

The incident mode-3 wave on the left is left-moving relative to the fluid and has negative norm. It is scattered into all four outgoing modes, including those right-moving relative to the fluid (modes 2 and 3 on the right). The outgoing mode 1 on the left and mode 2 on the right have positive norm, so conservation of norm again implies amplification of the incident wave. We denote the coefficient for scattering of the incident mode 3 into outgoing mode $n$ by $S_{n,3}$, The expansions of (\ref{3in}) into asymptotic waves then take the form
 \begin{align}   
\phi^{\text{3in}}(x) \stackrel{x\ll 0}{\sim}  & \, \phi_3^-(x) + S_{1,3} \phi_1^-(x) + S_{4,3} \phi_4^-(x), \label{3inleft} \\
\phi^{\text{3in}}(x) \stackrel{x\gg 0}{\sim}  & \,   S_{2,3} \phi_2^+(x)  +  S_{3,3} \phi_3^+(x), \label{3inright}
\end{align}
with scattering coefficients that follow from (\ref{H2left}), (\ref{H2right}), (\ref{F2left}), (\ref{F2right}) and (\ref{3in}). We calculate the absolute values of the scattering coefficients and find expressions that already appear in (\ref{S12})--(\ref{S42}):
 \begin{gather}   
 \left| S_{1,3} \right|^2    =   \left| S_{3,2} \right|^2 , \qquad   \left| S_{2,3} \right|^2    =   \left| S_{4,2} \right|^2 ,    \label{S2S3a}  \\
 \left| S_{3,3} \right|^2    =   \left| S_{1,2} \right|^2 , \qquad   \left| S_{4,3} \right|^2    =   \left| S_{2,2} \right|^2 .      \label{S2S3b} 
\end{gather}
The scattering amplitudes for mode 3 obey the identity that follows from constancy of the norm flux:
\begin{equation}
-1-  \left| S_{1,3} \right|^2  +   \left| S_{4,3} \right|^2 =  \left| S_{2,3} \right|^2  -  \left| S_{3,3} \right|^2 .
\end{equation}
There is also a simple relation analogous to (\ref{ratio}):
\begin{equation}
\frac{  \left| S_{1,3} \right|  }{  \left| S_{4,3} \right| } = \frac{  \left| S_{3,3} \right|  }{  \left| S_{2,3} \right| }  = e^{ -\frac{\pi\omega}{\alpha} } .   \label{ratio2}
\end{equation}

By means of (\ref{S12as})--(\ref{S42as}) we find immediately the leading-order scattering amplitudes (\ref{S2S3a}) and (\ref{S2S3b}) as $k_c\to\infty$. The scattering amplitudes thus reproduce the non-dispersive Hawking result
\begin{gather}
 \left| S_{2,3} \right|^2    =   \left| S_{3,3} \right|^2  = 0 ,  \label{S23haw}  \\
\left| S_{4,3} \right|^2    =  \left(  1 - e^{ -\frac{2\pi\omega}{\alpha} } \right)^{-1} ,    \quad
    \left| S_{1,3} \right|^2    =   \left(  e^{ \frac{2\pi\omega}{\alpha} } -1  \right)^{-1} ,    \label{S43haw}   
\end{gather}
in which there is no scattering into modes right-moving relative to the fluid and $\left| S_{1,3} \right|^2$ has the form of the Planck distribution.

The scattering of the incident mode 3 into modes with opposite norm again implies spontaneous emission in the quantum theory~\cite{rob12,bro95,unr95}. In the black-hole binary, the expectation value for the number of quanta in mode 3 emitted to the left (reading Fig.~\ref{fig:3in} from top to bottom) is $ \left| S_{1,3} \right|^2 + \left| S_{2,3} \right|^2 $. From (\ref{S2S3a}) we see that this is equal to the result we obtained for the number of quanta in mode 2  emitted to the left by the black-hole binary. The symmetry of the problem shows that this is also the number of quanta in the low-$k$ modes 1 and 4 emitted to the right (see Fig.~\ref{fig:rays}). There is thus emission of the same non-thermal spectrum of radiation in all the modes, both left-moving and right-moving relative to the fluid.

Our results were derived for the strictly linear flow profile (\ref{flow}), but our analysis showed that there is no wave scattering in the far-left and far-right regions of the flow. The scattering coefficients will therefore be the same for a flow profile that flattens out far from the horizons, provided the change in the flow velocity with distance is slow enough not to induce further scattering.

\section{Concluding remarks} 
We chose the fourth-order wave equation (\ref{wave}) because it is relatively simple while still being applicable to a physical system (the flowing BEC). The same equation with $k_c\to ik_c$ has dispersion that is normal rather than anomalous, but this gives a fourth-order normal dispersion relation $\omega(k)$ that is not monotonic in $k$ in the fluid frame. A monotonic normal dispersion relation leads to singular wave propagation in the linear flow profile, as some modes are infinitely blue-shifted as they approach any point where the flow speed is zero~\cite{bar05}. In reality such modes would be heavily damped as their wavelengths go to zero. For these reasons we have not treated the case of normal dispersion in the linear flow profile.

Our motivation was to obtain an exact solution for the Hawking effect in the presence of dispersion. The linear flow profile has two horizons but it can be solved exactly and the scattering amplitudes (\ref{S12})--(\ref{S42}) and (\ref{S2S3a})--(\ref{S2S3b}) are our final results. They demonstrate in exact formulas how dispersion changes the Hawking effect in one particular flow profile. 

For the wave equation (\ref{wave}), the new qualitative feature introduced by the dispersive term is the reflection of waves, i.e.\  the scattering of right-movers relative to the fluid into left-movers and vice versa. In the absence of dispersion there is no reflection because equation (\ref{wave}) is then exactly the 1+1-dimensional wave equation in curved space-time, and conformal flatness of the metric tensor leads to a strict separation of left- and right-movers. Dispersion causes coupling between the left- and right-movers and this is why reflection occurs in our example. The modification of the Hawking effect due to dispersion is related to the amount of refection because the total scattering into all channels must conserve the norm. The precise relationship between the various scattering channels will depend on the flow profile, even when the dispersion is fixed. For more complicated flow functions $v(x)$ than the one considered here, the scattering coefficients for the wave equation (\ref{wave}) will also be more complicated, if indeed exact results can be found.

Instead of trying to solve the wave scattering in a given flow profile, an alternative possibility is to design profiles that give a desired scattering. This approach has been fruitfully pursued in optics and quantum mechanics~\cite{ber90,lek07,hor15,lon15,phi16,hor16a,hor16b,lon16}. An important lesson from this work is that a breakdown of the geometrical-optics approximation does not necessarily imply scattering. In fact several infinite classes of inhomogeneous profiles are known in optics that have strictly zero scattering, even when the geometrical-optics approximation is violated arbitrarily badly. If these techniques can be extended to the wave equation in a moving medium, then dramatic differences in the spectrum of spontaneous quantum emission may be achieved by careful control of the flow profile.

\section*{Acknowledgements}
I am indebted to R.\ Parentani for informing me of the close connection between this work and ref.~\cite{bus12}, and also for commenting on the manuscript. I also thank S.A.R.\ Horsley, C.\ G.\ King and R.\ J.\ Churchill for many helpful discussions.

\appendix
\section{Conserved quantities}  \label{app:conserved}
An action giving the  wave equation (\ref{wave}) can be written in a general form that allows for arbitrary dispersion~\cite{unr95}:
\begin{align}
S=\int \int dt \, dx\left[\frac{1}{2}(\partial_t\psi^*+v\partial_x\psi^*) (\partial_t\psi+v\partial_x\psi) \right. \nonumber \\
\left.  -\frac{1}{2}F^*(-i\partial_x)\psi^*F(-i\partial_x)\psi \right],     \label{act}   \\
F(-i\partial_x)=\sum_{n=0}^\infty (-1)^{n+1} i b_{2n+1} \partial_x^{2n+1}. \qquad
\end{align}
This gives the wave equation
\begin{equation}  \label{wavegen}
\partial_t (\partial_t+v\partial_x )\psi+\partial_x (v\partial_t+v^2\partial_x )\psi+F^2(-i\partial_x)\psi=0,
\end{equation}
with the dispersion relation
\begin{equation} \label{disprelgen}
(\omega-vk)^2=F^2(k).
\end{equation}
The general dispersive equation (\ref{wavegen}) has spatial derivatives of $\psi$ of all even orders (terms in the wave equation with an odd number of derivatives would give dissipation). The fourth-order equation (\ref{wave}) corresponds to $F^2(-i\partial_x)=-\partial_x^2+\frac{1}{k_c^2}\partial_x^4$, which gives an $F(-i\partial_x)$ that is defined by the power series
\begin{equation}   \label{F} 
F(k)=k\sqrt{1+\frac{k^2}{k_c^2}}=\sum_{n=0}^\infty   \binom{\frac{1}{2}}{n} \frac{k^{2n+1}}{k_c^{2n}}.
\end{equation}
In (\ref{act}) we allow $\psi(x,t)$ to be complex to see better the quantities conserved by (\ref{wavegen}). We derive the conservation laws for arbitrary dispersion and then specialise to the fourth-order equation (\ref{wave}).

The action (\ref{act}) is invariant under the $U(1)$ transformation $\psi\to e^{i\theta}\psi$ and also under time translation (since $v(x)$ is time independent). The conserved quantities associated with these symmetries are the norm and the pseudo-energy, respectively. To construct the conservation laws we must apply Noether's theorem to an action with an (in general) unbounded number of terms containing derivatives of arbitrarily high order. The method for applying Noether's theorem to such actions is described in~\cite{phi11}, with examples from dispersive optics. We refer to~\cite{phi11} for the technicalities of how to construct the conservation laws and here quote the results for the norm and pseudo-energy. The norm density $\rho_N(x,t)$ and norm flux $s_N(x,t)$ are
\begin{align}
\rho_N   = \, & i\psi^*(\partial_t\psi+v\partial_x\psi) +\text{c.c.},   \label{rhoN}  \\
s_N  = \, & iv\psi^*(\partial_t\psi+v\partial_x\psi)    \nonumber  \\
  & +\sum_{n=0}^\infty \sum_{m=0}^{2n} (-1)^{n+m} b_{2n+1}[\partial_x^mF(-i\partial_x)\psi ]\partial_x^{2n-m}\psi^*     \nonumber \\
  & +\text{c.c.},    \label{sN}
\end{align}
where c.c.\ means complex conjugate. It is straightforward to verify that the norm conservation law
\begin{equation}   \label{conlaw}
\partial_t \rho_N (x,t)+ \partial_x s_N (x,t)=0
\end{equation}
holds for waves satisfying the general dispersive equation (\ref{wavegen}). The pseudo-energy density $\rho_E(x,t)$ and pseudo-energy flux $s_E(x,t)$ are
\begin{align}
\rho_E  = \, &\frac{1}{2}\left( \partial_t\psi^* \partial_t\psi - v^2\partial_x\psi^* \partial_x\psi \right)  \nonumber  \\ 
   &+\frac{1}{2} F^*(-i\partial_x)\psi^* F(-i\partial_x)\psi ,   \label{rhoE}  \\
s_E  = \, & \frac{1}{2} v \partial_t\psi^*(\partial_t\psi+v\partial_x\psi)    \nonumber  \\
   - & \frac{i}{2} \sum_{n=0}^\infty \sum_{m=0}^{2n} (-1)^{n+m} b_{2n+1}[\partial_x^mF(-i\partial_x)\psi ]\partial_x^{2n-m}\partial_t\psi^*     \nonumber \\
  & +\text{c.c.}    \label{sE}
\end{align}
These also obey the conservation law of form (\ref{conlaw}), because of (\ref{wavegen}).

For monochromatic waves $\psi(x,t)=e^{-i\omega t} \phi(x)$, the densities (\ref{rhoN}) and (\ref{rhoE}), and fluxes (\ref{sN}) and (\ref{sE}), are clearly time-independent. The conservation law (\ref{conlaw}) thus gives for monochromatic waves
\begin{equation}  \label{sNconst}
\partial_x s_N =0,
\end{equation}
so that the flux is the same at each point in the inhomogeneous flow. This constant-flux condition is of great importance in analysing wave propagation in the fluid. The fourth-order wave equation (\ref{wave}) corresponds to $F$ given by (\ref{F}), and in this case the norm flux (\ref{sN}) reduces to a finite number of terms:
\begin{align}
s_N (x,t) =  \, &    iv\psi^*(\partial_t\psi+v\partial_x\psi)   -i  \psi^* \partial_x \psi     \nonumber  \\
&  + i k_c^{-2} \left(  \psi^* \partial_x^3 \psi   - \partial_x \psi^* \partial_x^2 \psi   \right)  +\text{c.c.}.        \label{sN4th} 
\end{align}
The constancy of this flux for monochromatic waves will be referred to throughout.

\section{Asymptotics of the wave equation}  \label{app:asymptotic}
The main aim here is to understand the wave equation (\ref{wave}) in the asymptotic regions $|x|\to\infty$ of the linear flow profile. We show that waves in the linear profile must reduce, as $|x|\to\infty$, to non-interacting wave components associated with the dispersion relation. The norm flux (\ref{sN4th}) for these wave components is then calculated.

The issue addressed here could be phrased as that of finding the WKB solutions of the wave equation in the flow. But it is worth pointing out some significant differences between asymptotic solutions for sound in a flowing fluid and the familiar WKB solutions in optics and quantum mechanics. For the Helmholtz equation (equivalently the time-independent Schr\"{o}dinger equation) in an inhomogeneous medium, there is a simple criterion for the WKB solutions to be good approximations, namely that the fractional change in the refractive index must be very small over a local wavelength. This criterion is clearly satisfied as $|x|\to\infty$ if the permittivity profile approaches constant values. But the Helmholtz equation has the property that the WKB criterion is satisfied for large $|x|$ even for profiles that diverge as $|x|\to\infty$, because the local wavelength goes to zero. This means that general WKB solutions can be written which are functionals of an arbitrary permittivity profile and these will always be valid as $|x|\to\infty$.  The wave equation for sound in a moving fluid shows important differences. One can compute leading-order asymptotic solutions in this case also, but the criterion for their validity is not very simple. For all flow profiles with regions where $v'(x)$ is very small, leading-order asymptotic solutions that are accurate in these regions can be derived as functionals of $v(x)$. These functionals, however, do \emph{not} give the leading-order asymptotic solutions for the linear profile in the regions $|x|\to\infty$. This is because $v'(x)$ stays constant as $|x|\to\infty$ for the linear profile while the wavelengths of some roots of the dispersion relation get larger.

For completeness, we first give the leading-order asymptotic solutions valid in any regions where $v'(x)$ is ``small" and allowing for arbitrary dispersion. Then we treat the leading-order asymptotic solutions of (\ref{wave}) for the linear profile (\ref{flow}), in the regions $|x|\to\infty$.

\subsection{Asymptotics in regions of slowly varying flow velocity}  
 The wave equation (\ref{wavegen}) in the case of arbitrary dispersion gives the monochromatic equation
\begin{equation}  \label{monogen}
 \left[\omega^2  +i\omega v'+2 v (i\omega-v')\partial_x   -v^2 \partial_x^2 -F^2(-i\partial_x)   \right]\phi=0.
\end{equation}
Following a standard approach to the WKB approximation in quantum mechanics~\cite{merz}, we substitute $\phi(x)= e^{i\chi(x)}$ and arrange (\ref{monogen}) as
\begin{align}  
(\omega-v\chi')^2   - & F^2(\chi')  =   -i\omega v'+2ivv'\chi'+iv^2\chi''     \nonumber  \\
&+e^{-i\chi}\left[F^2(-i\partial_x) -F^2(\chi')    \right] e^{i\chi}.  \label{wkb1}
\end{align}
In regions where $v'$, and therefore $\chi'$, are nearly constant, terms on the right-hand side of (\ref{wkb1}) are small compared to terms on the left-hand side. We therefore iterate (\ref{wkb1}) as follows. To lowest order the solution of (\ref{wkb1}) is $\chi_0'$, satisfying
\begin{equation}    \label{wkb2}
(\omega-v\chi_0')^2   -  F^2(\chi_0')  = 0,
\end{equation}
i.e.\ the branches of the dispersion relation (\ref{disprelgen}). The lowest order correction $\chi_1'$ to $\chi_0'$ is found by inserting $\chi'=\chi_0'+\chi_1'$ into (\ref{wkb1}), applying (\ref{wkb2}), and keeping only terms linear in small quantities of the same order as $\chi_1'$, i.e.\ $\chi_0''$, $\chi_1'$ and $v'$. This gives
\begin{align}  
-2(\omega-v\chi_0')v\chi_1'   - & 2F(\chi_0')F'(\chi_0')\chi_1'  =   -i\omega v'+2ivv'\chi_0'    \nonumber  \\
&  +iv^2\chi_0''  -i \frac{d}{dx}\left[F(\chi_0') F'(\chi_0') \right] ,  \label{wkb3}
\end{align}
which yields the following solution for $\chi_1'$:
\begin{align}
\chi_1'  &=\frac{i}{2}\frac{d}{dx} \ln \left|(\omega-v\chi_0')v+F(\chi_0') F'(\chi_0') \right|   \nonumber  \\
&= \frac{i}{2}\frac{d}{dx} \ln \left|F(\chi_0') V_g(\chi_0') \right|.
\end{align}
The last expression contains $V_g(\chi_0')$, the group velocity in the laboratory frame of the mode given by the root $\chi_0'$ of the dispersion relation (\ref{wkb2}), i.e.\ 
\begin{equation}
V_g(k)=v\pm F'(k),
\end{equation}
where the sign depends on the branch $k$. The solution for $\phi(x)= e^{i\chi(x)}$ to order $\chi_1'$ is thus
\begin{equation}  \label{wkbsoln}
\phi(x)\sim  \frac{1}{2\sqrt{\left|F(\chi_0') V_g(\chi_0') \right|}} e^{i\chi_0},
\end{equation}
where a normalization factor of $1/2$ is inserted. This result, valid for arbitrary dispersion, can also be derived from an analysis of the wave equation  (\ref{monogen}) in $k$-space~\cite{rob12}. 

The norm density (\ref{rhoN})  and pseudo-energy density (\ref{rhoE}) for the asymptotic solutions (\ref{wkbsoln}) is positive or negative according to the sign of the co-moving frequency $\omega-v\chi_0'$. Norm and pseudo-energy are transported at the group velocity of the mode, i.e.\ $s_N/\rho_N=s_E/\rho_E=V_g(\chi_0')$. 

The norm flux (\ref{sN}) of the asymptotic solutions (\ref{wkbsoln}) is equal to $\pm1$ (to the same order of approximation). The sign of the norm flux is given by the product of two signs: the sign of the norm being transported by the mode (given by the sign of the co-moving frequency) and the sign of the mode's group velocity. A superposition of asymptotic solutions (\ref{wkbsoln}) for different roots $\chi_0'$ of the dispersion relation has the important property that its norm flux is just the sum of the fluxes of the individual components in the superposition, i.e.\ all cross terms in (\ref{sN}) involving different components cancel out. 

\subsection{Asymptotics of the wave equation in the linear profile}  
Here we confine attention to the fourth-order equation (\ref{mono}) in the linear profile (\ref{flow}), and find its asymptotic solutions in the regions $|x|\to\infty$. 

As noted in Sec.~\ref{sec:wave}, the four roots of the dispersion relation (\ref{disprel}) in the linear profile have complicated expressions, but here we require only their asymptotic expansions for large $|x|$. We denote the roots by $k_n(x)$, where $n$ labels the ray solutions 1 to 4 discussed in  Sec.~\ref{sec:wave}.  For large positive $x$ the first few terms of the expansions of  $k_n(x)$ are
\begin{align}  
k_1(x) & \sim \frac{\omega}{\alpha x}\left( -1+\frac{1}{\alpha x} -\frac{1}{\alpha^2 x^2} +\frac{2k_c^2+\omega^2}{2k_c^2\alpha^3 x^3} \right) ,    \label{k1}  \\
k_2(x)  &  \sim \alpha k_c x - \frac{k_c-2\omega}{2\alpha x}  - \frac{k_c^2-8k_c\omega+8\omega^2}{8k_c\alpha^3 x^3} , \label{k2}  \\
k_3(x)  &  \sim -\alpha k_c x + \frac{k_c+2\omega}{2\alpha x}  + \frac{k_c^2+8k_c\omega+8\omega^2}{8k_c\alpha^3 x^3},   \label{k3}  \\
k_4(x)  &  \sim \frac{\omega}{\alpha x}\left( -1-\frac{1}{\alpha x} -\frac{1}{\alpha^2 x^2} -\frac{2k_c^2+\omega^2}{2k_c^2\alpha^3 x^3} \right) .    \label{k4}  
\end{align}

The invariance of the dispersion relation under $x\to-x,\ k\to-k$ leads to the following (exact) relations: $k_1(-x)=-k_2(x)$,  $k_3(-x)=-k_4(x)$ (this can be seen in the dispersion plots in Fig.~\ref{fig:disp}). We can thus easily obtain from (\ref{k1})--(\ref{k4}) the asymptotic expansions of the wave-vectors $k_n(x)$ for large negative $x$. 

We first seek asymptotic solutions to (\ref{mono}) for large positive $x$. In view of (\ref{wkbsoln}), we make the Ansatz
\begin{equation}  \label{asym1}
\phi^+_n(x)\sim A^+_n(x) \exp\left[i\int^x dx\, k_n(x)\right],
\end{equation}
where $A^+_n(x)$ are unknown amplitude functions for the modes 1 to 4, and the superscript $+$ labels the region (large positive $x$) in which the expansion (\ref{asym1}) is to be valid. We insert (\ref{asym1}) for $n=1,\dots,4$ into the wave equation (\ref{mono}) and demand that it be satisfied for $x\to\infty$. For modes 1 and 4 the wave equation is satisfied by (\ref{asym1}) as $x\to\infty$ even with constant $A^+_1$ and   $A^+_4$. By demanding that  $A^+_1(x)$ and   $A^+_4(x)$ increase the accuracy of the asymptotic solutions (\ref{asym1}), so that the wave equation is satisfied to higher orders of $1/x$, we can build up the required amplitudes $A^+_1(x)$ and  $A^+_4(x)$ as  asymptotic series. For modes 2 and 3 we perform the same procedure, but here the wave equation is not satisfied by (\ref{asym1}) to any order of $1/x$ without $x$-dependent amplitudes $A^+_2(x)$ and   $A^+_3(x)$. The amplitudes $A^+_n$, to the orders consistent with the accuracy of the expansions (\ref{k1})--(\ref{k4}), are found to be
\begin{align}  
A^+_1(x) & \sim \frac{1}{\sqrt{2\omega}}\left( 1+\frac{i\omega^3}{6k_c^2\alpha^4 x^3}  \right) ,    \label{A1}  \\
A^+_2(x)  &  \sim  \frac{1}{\sqrt{2k_c\alpha^3x^3}}\left( 1+ \frac{12 k_c-27i\alpha-32 \omega}{16k_c\alpha^2 x^2}  \right)  , \label{A2}  \\
A^+_3(x) &  \sim   \frac{1}{\sqrt{2k_c\alpha^3x^3}}\left( 1+ \frac{12 k_c+27i\alpha+32 \omega}{16k_c\alpha^2 x^2}  \right)  , \label{A3}  \\
A^+_4(x) &  \sim  \frac{1}{\sqrt{2\omega}}\left( 1-\frac{i\omega^3}{6k_c^2\alpha^4 x^3}  \right) .    \label{A4}  
\end{align}
where convenient constant normalization factors have been included. Using these and (\ref{k1})--(\ref{k4}) in (\ref{asym1}) we obtain asymptotic solutions for the four modes. We will only need the leading order of (\ref{asym1}) for modes 2 and 3,  but we require the first three orders for modes 1 and 4, as follows: 
\begin{align}  
\phi^+_1(x) & \sim \frac{ x^{-i\omega/\alpha}}{\sqrt{2\omega}}\left( 1-\frac{i\omega}{\alpha^2 x}+\frac{i\omega(\alpha+i\omega)}{2\alpha^4 x^2}  \right) ,    \label{+1}  \\
\phi^+_2(x)  &  \sim  \frac{ x^{-\frac{3}{2}+i\frac{2\omega-k_c}{2\alpha}}}{\sqrt{2k_c\alpha^3}} \exp\left( \frac{i \alpha k_c x^2}{2}  \right)  , \label{+2}  \\
\phi^+_3(x) &  \sim  \frac{ x^{-\frac{3}{2}+i\frac{2\omega+k_c}{2\alpha}}}{\sqrt{2k_c\alpha^3}} \exp\left(- \frac{i \alpha k_c x^2}{2}  \right) , \label{+3}  \\
\phi^+_4(x) &  \sim \frac{ x^{-i\omega/\alpha}}{\sqrt{2\omega}}\left( 1+\frac{i\omega}{\alpha^2 x}+\frac{i\omega(\alpha+i\omega)}{2\alpha^4 x^2}  \right) .    \label{+4}  
\end{align}

One can show that the amplitudes (\ref{A2}) and (\ref{A3}) for modes 2 and 3 are the same as would be obtained by using the result (\ref{wkbsoln}) for a slowly varying flow velocity. But the amplitudes (\ref{A1}) and (\ref{A4}) for modes 1 and 4 are not given correctly by (\ref{wkbsoln}). 

The four expressions (\ref{+1})--(\ref{+4}) are also asymptotic solutions for large negative $x$, but the identification of each with one of the four modes is different in the region $x\to-\infty$. It is straightforward to find  the corresponding mode in each case, and the four asymptotic solutions $\phi^-_n(x)$ for $x\to-\infty$ take the form
\begin{align}  
\phi^-_1(x) & \sim  \frac{ \left|x\right|^{-\frac{1}{2}+i\frac{2\omega-k_c}{2\alpha}}}{x\sqrt{2k_c\alpha^3}} \exp\left( \frac{ i \alpha k_c x^2}{2}   \right) ,    \label{-1}  \\
\phi^-_2(x)  &  \sim \frac{ \left|x\right|^{-i\omega/\alpha}}{\sqrt{2\omega}}\left( 1+\frac{i\omega}{\alpha^2 x}+\frac{i\omega(\alpha+i\omega)}{2\alpha^4 x^2}  \right) , \label{-2}  \\
\phi^-_3(x) &  \sim \frac{ \left|x\right|^{-i\omega/\alpha}}{\sqrt{2\omega}}\left( 1-\frac{i\omega}{\alpha^2 x}+\frac{i\omega(\alpha+i\omega)}{2\alpha^4 x^2}  \right)  , \label{-3}  \\
\phi^-_4(x) &  \sim \frac{ \left|x\right|^{-\frac{1}{2}+i\frac{2\omega+k_c}{2\alpha}}}{x\sqrt{2k_c\alpha^3}} \exp\left( - \frac{i \alpha k_c x^2}{2}  \right) .    \label{-4}  
\end{align}

The norm flux (\ref{sN}) of each of the asymptotic solutions (\ref{+1})--(\ref{-4}) is equal to $\pm1$, to leading order. Modes 2 and 4 have norm flux equal to $+1$ while modes 1 and 3  have norm flux of $-1$. The sign of the norm flux is the product of the sign of the norm (positive for modes 1 and 2, negative for modes 3 and 4) and the group velocity (positive for modes 2 and 3, negative for modes 1 and 4). A superposition of the asymptotic solutions has a norm flux that is the sum of the fluxes of the individual mode components, i.e.\ all cross terms in (\ref{sN}) between different modes cancel out, to leading order.



\begin{thebibliography}{99}

\bibitem{unr81}
W.\ G.\ Unruh, Phys.\ Rev.\ Lett.\  {\bf 46}, 1351 (1981).

\bibitem{bar05}
C. Barcelo, S.\ Liberati, and M.\ Visser, Living Rev.\ Rel.\ {\bf 8}, 12 (2005).

\bibitem{rob12}
S.\ J.\ Robertson, J.\ Phys.\ B: At.\ Mol.\ Opt.\ Phys.\ {\bf 45}, 163001 (2012).

\bibitem{haw74}
S.\ W.\ Hawking, Nature (London) {\bf 248}, 30 (1974).

\bibitem{bro95}
R.\ Brout, S.\ Massar, R.\ Parentani, and Ph.\ Spindel, Phys. Rep. {\bf 260}, 329 (1995).

\bibitem{Gardiner}
 C.\ W.\ Gardiner and P.\ Zoller, {\it Quantum Noise}, 3rd ed.\ (Springer, Berlin, 2010).
 
\bibitem{gar00}
L.\ J.\ Garay, J.\ R.\ Anglin, J.\ I.\ Cirac, and P.\ Zoller, Phys.\ Rev.\ Lett.\  {\bf 85}, 4643 (2000).

\bibitem{sch02}
R.\ Sch\"{u}tzhold and W.\ G.\ Unruh, Phys.\ Rev.\ D  {\bf 66}, 044019 (2002).

\bibitem{rou08}
G.\ Rousseaux, C.\ Mathis, P.\ Maissa, T.\ G.\ Philbin, and U.\ Leonhardt, New J.\ Phys.\ {\bf 10}, 053015 (2008).

\bibitem{phi08}
T.\ G.\ Philbin, C.\ Kuklewicz, S.\ Robertson, S.\ Hill, F.\ K\"onig, and U.\ Leonhardt, Science {\bf 319}, 1367 (2008).

\bibitem{bel10}
F.\ Belgiorno, S.\ L.\ Cacciatori, G.\ Ortenzi, V.\ G.\ Sala, and D.\ Faccio, Phys.\ Rev.\ Lett.\  {\bf 104}, 140403 (2010).

\bibitem{lah10}
O.\ Lahav, A.\ Itah, A.\ Blumkin, C.\ Gordon, S.\ Rinott, A.\ Zayats, and J.\ Steinhauer, Phys.\ Rev.\ Lett.\  {\bf 105}, 240401 (2010).

\bibitem{wei11}
S.\ Weinfurtner, E.\ W.\ Tedford, M.\ C.\ J.\ Penrice, W.\ G.\ Unruh, and G.\ A.\ Lawrence,  Phys.\ Rev.\ Lett.\  {\bf 106}, 021302 (2011).

\bibitem{ste14}
J.\ Steinhauer, Nat.\ Phys.\  {\bf 10}, 864 (2014).

\bibitem{ngu15}
H.\ S.\ Nguyen, {\it et al}, Phys.\ Rev.\ Lett.\ {\bf 114}, 036402 (2015).

\bibitem{ste15}
J.\ Steinhauer, Nat. Phys.  http://dx.doi.org/10.1038/nphys3863 (2016).

\bibitem{euv15a}
L.-P.\ Euv\'{e}, F.\ Michel, R.\ Parentani, T.G. Philbin and G.\ Rousseaux, Phys. Rev. Lett. {\bf 117}, 121301 (2016).

\bibitem{Jacobson}
T.\ Jacobson, Phys.\ Rev.\ D {\bf 44}, 1731 (1991); {\bf 48}, 728 (1993).

\bibitem{unr95}
W.\ G.\ Unruh, Phys.\ Rev.\ D  {\bf 51}, 2827 (1995).

\bibitem{cor96}
S.\ Corley and T.\ Jacobson, Phys.\ Rev.\ D {\bf 54}, 1568 (1996).

\bibitem{mac09}
J.\ Macher and R.\ Parentani, Phys.\ Rev.\ D  {\bf 79}, 124008 (2009).

\bibitem{leo12}
U.\ Leonhardt and S.\ Robertson, New J.\ Phys.\  {\bf 14}, 053003 (2012).

\bibitem{fin12}
S.\ Finazzi and R.\ Parentani, Phys.\ Rev.\ D  {\bf 85}, 124027 (2012).

\bibitem{cou12}
A.\ Coutant, R.\ Parentani, and S.\ Finazzi, Phys.\ Rev.\ D  {\bf 85}, 024021 (2012).

\bibitem{mic14}
F.\ Michel and R.\ Parentani, Phys.\ Rev.\ D  {\bf 90}, 044033 (2014).

\bibitem{rob14}
S.\ Robertson and U.\ Leonhardt, Phys.\ Rev.\ E  {\bf 90}, 053302 (2014); {\bf 90}, 053303 (2014).

\bibitem{euv15}
L.-P.\ Euv\'{e}, F.\ Michel, R.\ Parentani and G.\ Rousseaux, Phys.\ Rev.\ D  {\bf 91}, 024020 (2015).

\bibitem{rou10}
G.\ Rousseaux, P.\ Ma\"{i}ssa, C.\ Mathis, P.\ Coullet, T.\ G.\ Philbin, and U.\ Leonhardt, New J. Phys. {\bf 12}, 095018
(2010).

\bibitem{pel15}
C. Peloquin, L.-P.\ Euv\'{e}, T.G. Philbin and G.\ Rousseaux, Phys.\ Rev.\ D  {\bf 93}, 084032 (2016).

\bibitem{bro95a}
R.\ Brout, S.\ Massar, R.\ Parentani, and Ph.\ Spindel, Phys.\ Rev.\ D {\bf 52}, 4559 (1995).

\bibitem{cor98}
S.\ Corley, Phys.\ Rev.\ D {\bf 57}, 6280 (1998).

\bibitem{him00}
Y.\ Himemoto and T.\ Tanaka, Phys.\ Rev.\ D {\bf 61}, 064004 (2000).

\bibitem{sai00}
H.\ Saida and M.\ Sakagami, Phys.\ Rev.\ D {\bf 61}, 084023 (2000).

\bibitem{unr05}
W.\ G.\ Unruh and R.\ Sch\"{u}tzhold, Phys.\ Rev.\ D  {\bf 71}, 024028 (2005).

\bibitem{mac09b}
J.\ Macher and R.\ Parentani,  Phys.\ Rev.\ A  {\bf 80}, 043601 (2009).

\bibitem{cou16}
A.\ Coutant and S.\ Weinfurtner, Phys. Rev. D {\bf 94}, 064026 (2016).

\bibitem{rob16}
S.\ Robertson, F.\ Michel and R.\ Parentani, Phys. Rev. D {\bf 93}, 124060 (2016).

\bibitem{mic16}
F.\ Michel and R.\ Parentani,  arXiv:1605.09752[cond-mat.quant-gas].

\bibitem{bus12}
X.\ Busch and R.\ Parentani, Phys.\ Rev.\ D {\bf 86}, 104033 (2012).

\bibitem{sch08}
R.\ Sch\"{u}tzhold and W.\ G.\ Unruh, Phys.\ Rev.\ D  {\bf 78}, 041504(R) (2008).

\bibitem{heading}
J.\ Heading, {\it An Introduction to Phase Integral Methods}, (Dover, New York, 2013).

\bibitem{dlmf}
NIST Digital Library of Mathematical Functions. http://dlmf.nist.gov/.

\bibitem{titchmarsch}
E.\ C.\ Titchmarsch, {\it Introduction to the Theory of Fourier Integrals}, 2nd ed.\ (Oxford University Press, Oxford, 1948).

\bibitem{ablowitz}
M.\ J.\  Ablowitz and A.\ S.\ Fokas, {\it Complex Variables}, 2nd ed.\ (Cambridge University Press, Cambridge, 2003).

\bibitem{whittaker}
E.\ T.\  Whittaker and G.\ N.\ Watson, {\it A Course of Modern Analysis}, 4th ed.\ (Cambridge University Press, Cambridge, 1927).

\bibitem{htf1}
A.\  Erd\'{e}lyi, F.\ Oberhettinger and F.\ G.\ Tricomi, {\it Higher Transcendental Functions}, Vol.\ I (McGraw-Hill, New York, 1953).

\bibitem{ada13}
J.\ Adamek, X.\ Busch and R.\ Parentani, Phys.\ Rev.\ D  {\bf 87}, 124039 (2013).

\bibitem{rob15}
S.\ Robertson and R.\ Parentani, Phys.\ Rev.\ D  {\bf 92}, 044043 (2015).

\bibitem{ber90}
M.\ V.\ Berry and C.\ J.\ Howels, J.\ Phys.\ A: Math.\ Gen.\ {\bf 23}, L243 (1990).

\bibitem{lek07}
J.\ Lekner, Am.\ J.\ Phys.\ {\bf 75}, 1151 (2007).

\bibitem{hor15}
S.\ A.\ R.\ Horsley, M.\ Artoni and G.\ C.\ La Rocca, Nat.\ Photonics {\bf 9}, 436 (2015).

\bibitem{lon15}
S.\ Longhi, Europhys.\ Lett.\ {\bf 112}, 64001 (2015).

\bibitem{phi16}
T.\ G.\ Philbin,  J.\ Opt.\ {\bf 18}, 01LT01 (2016).

\bibitem{hor16a}
S.\ A.\ R.\ Horsley, C.\ G.\ King and T.\ G.\ Philbin,  J.\ Opt.\ {\bf 18}, 044016 (2016).

\bibitem{hor16b}
S.\ A.\ R.\ Horsley, J. Opt. {\bf 18}, 085104 (2016).

\bibitem{lon16}
S.\ Longhi, Opt. Lett. {\bf 41}, 3727 (2016).

\bibitem{phi11}
T.\ G.\ Philbin, Phys.\ Rev.\ A {\bf 83}, 013823 (2011); {\bf 85}, 059902(E) (2012);
T.\ G.\ Philbin and O.\ Allanson, Phys.\ Rev.\ A {\bf 86}, 055802 (2012).

\bibitem{merz}
E.\ Merzbacher, {\it Quantum Mechanics}, 3rd ed.\ (Wiley, New York, 1998).

\end{thebibliography}
\end{document}